\documentclass[sigconf,10pt,nonacm,screen]{acmart}
\renewcommand\footnotetextcopyrightpermission[1]{}
\setcopyright{none}
\acmDOI{}
\acmISBN{}

\usepackage[english]{babel}
\usepackage{graphicx}
\usepackage{subcaption}
\usepackage{booktabs}
\usepackage{longtable}
\usepackage{listings}
\usepackage{amsmath}
\usepackage{comment}
\usepackage{array}
\usepackage[table]{xcolor}
\usepackage{soul}
\usepackage[normalem]{ulem}
\usepackage{algorithm}
\usepackage{algpseudocode}
\usepackage{geometry}

\usepackage{amssymb,amsthm,mathtools}
\usepackage{enumitem}
\usepackage{hyperref}
\usepackage{xspace}
\usepackage{tikz}

\newcolumntype{P}[1]{>{\centering\arraybackslash}p{#1}}
\newcommand{\squishlist}{
    \begin{list}{$\bullet$}{
        \setlength{\itemsep}{0pt}
        \setlength{\parsep}{2pt}
        \setlength{\topsep}{2pt}
        \setlength{\partopsep}{0pt}
        \setlength{\leftmargin}{1em}
        \setlength{\labelwidth}{0.5em}
        \setlength{\labelsep}{0.5em}
    }
}
\newcommand{\squishend}{
    \end{list}
}

\usepackage[most]{tcolorbox}
\newtcolorbox{promptbox}{
  colback=gray!5,
  colframe=black!70,
  boxrule=0.8pt,
  arc=4pt,
  width=\linewidth,
  breakable,
  fontupper=\footnotesize,
  left=2pt,
  right=2pt,
  top=2pt,
  bottom=2pt,
}

\begin{document}

\newcommand{\name}{IteRate\xspace}

\newcommand{\nWorkflows}{5\xspace}

\newcommand{\webImproveOE}{52\%}
\newcommand{\webImproveMinstrel}{21\%}

\newcommand{\videoImproveOE}{12\%}
\newcommand{\videoImproveMinstrel}{7\%}

\newcommand{\satImproveOE}{10\%}
\newcommand{\satImproveMinstrel}{21\%}

\newcommand{\webFCTOur}{2.97\,s}
\newcommand{\webFCTOE}{6.19\,s}
\newcommand{\webFCTMinstrel}{3.76\,s}

\newcommand{\satThroughputOur}{8.91\,Mbps}
\newcommand{\satThroughputOE}{8.08\,Mbps}
\newcommand{\satThroughputMinstrel}{7.35\,Mbps}

\newcommand{\videoMOSOur}{3.06}
\newcommand{\videoMOSOE}{2.73}
\newcommand{\videoMOSMinstrel}{2.86}
\newcommand{\aff}{
  \affiliation{
    \institution{MIT Computer Science and Artificial Intelligence Lab}
    \city{Cambridge}
    \state{Massachusetts}
    \country{USA}
  }
}

\title{\name: Autonomous AI Synthesis of In-Kernel eBPF Wi-Fi Rate Control Algorithms}
\author{James Lynch, Ziqian Liu, Snehadeep Gayen, Om Chabra, Hari Balakrishnan}
\aff
\email{{jclynch, z229liu, sgayen, omchabra, hari}@mit.edu}
\date{March 2026}

\renewcommand{\shortauthors}{Lynch et al.}


\begin{abstract}

Wi-Fi rate adaptation remains a persistent challenge in wireless networking. Deployed algorithms like Minstrel-HT have remained largely stagnant for over a decade, relying on hand-tuned heuristics that fail to generalize to the complexity of modern wireless environments. We present \name, an autonomous research system that closes the loop on rate control development. \name uses a multi-agent AI architecture to conduct the full scientific cycle: formulating hypotheses, writing eBPF programs that run inside the Linux kernel, deploying them over-the-air to  Wi-Fi devices, collecting fine-grained telemetry for analysis, and iterating based on experimental evidence, all without human intervention. IteRate makes three contributions. (1) a novel kernel module that exposes per-frame hardware telemetry including modulation and coding schemes (MCS) and retry counts to eBPF programs,
(2) a structured agentic AI architecture employing specialized agents for algorithm design, experiment execution, and data analysis, coordinated via a hypothesis-driven research protocol with persistent knowledge, and (3) a closed-loop pipeline that automates the cross-compilation, deployment, and evaluation of in-kernel logic onto embedded Wi-Fi targets. 

On a 58-node testbed running five workloads. relative to the well-known Minstrel algorithm, \name{} achieves \webImproveMinstrel{} faster web-page loads, \videoImproveMinstrel{} higher video quality of experience (QoE), and \satImproveMinstrel{} higher peak throughput. 
Our work demonstrates that AI agents, when equipped with appropriate kernel-level hooks and a disciplined scientific workflow, can effectively automate the research required to design Wi-Fi rate controllers.

\end{abstract}

\maketitle

\section{Introduction}
We introduce \name, the first system to enable AI-driven, online exploration of Wi-Fi bitrate selection algorithms directly on commodity hardware. Rather than relying on expert hand-crafted heuristics, \name operates as an embedded researcher, orchestrating on-device experiments, analyzing real-time telemetry, and learning from traces to synthesize new algorithms that adapt to prevailing conditions.

Bitrate selection runs at the MAC layer at the sender, selecting a modulation and coding scheme (MCS) for each transmitted frame. Because of multipath fading, client mobility, and transient interference, the ideal MCS varies with time~\cite{bicket2005bitrate,wong2006robust}. Suboptimal rate choices lead to degraded user experience, increased latency, and spectrum under-utilization. An ideal algorithm would adapt to changing conditions and also to application requirements.

Historically, bitrate selection has relied on expert-designed heuristics like Minstrel \cite{xia2013minstrel} and SampleRate \cite{bicket2005bitrate}. While these algorithms are computationally lightweight and fit easily within the strict timing constraints of MAC-layer firmware, they apply rigid, one-size-fits-all logic that fails to adapt to diverse channel dynamics and changing workloads (\S\ref{sec:evaluation}). In response, recent research has turned to machine learning (ML) and reinforcement learning (RL)~\cite{karmakar2017smartla,khastoo2020neura,queiros2022dara}. These ML approaches require heavy offline retraining, are opaque to expert engineers, and often violate the microsecond-level execution constraints of the MAC layer \cite{khastoo2020neura,yin2024adrx}.

In this paper, we ask: \emph{can we use large language models (LLMs) to synthesize practical bitrate algorithms tailored to the prevailing environment and workload?} Unlike traditional black-box ML models, LLMs can generate interpretable source code~\cite{glia}. This difference allows us to combine the adaptability of data-driven methods with the microsecond-level execution speeds of classic compiled heuristics. By leveraging LLMs to write and mutate MAC-layer logic, we can plausibly generate bespoke algorithms on the fly. 

While recent research has applied LLM-driven pipelines to some problems in networked systems (e.g., OpenEvolve \cite{openevolve}, Glia \cite{glia}, ADRS \cite{ADRS}), these frameworks assume static, repeatable environments where simulators are accurate and network states are predictable. Wireless networks violate these assumptions. Real-world RF environments are inherently chaotic, and even advanced simulators struggle to capture accurate physical-layer dynamics or hardware interaction with application workloads \cite{abedi2016tsimn,vutukuru2009softrate}. Hence, algorithms synthesized in simulation environments tend to overfit to idealized conditions and often perform poorly on real hardware~\cite{abedi2014trate,abedi2016tsimn}.

\name enables the online synthesis and evaluation of Wi-Fi rate control algorithms within a live network environment. Rather than using offline simulators, \name provides a programmable experimentation environment where newly generated algorithms are rapidly deployed, executed on the packet transmission path, and evaluated using real wireless traffic by AI agents to produce new candidate algorithms.

\name decouples high-level algorithm synthesis from low-level execution. An LLM-driven agent system operates as an autonomous researcher that observes real-time network telemetry, proposes candidate rate-control policies, and iteratively refines them based on empirical feedback. \name then transforms these policies into executable code that is safely injected into the live datapath, enabling microsecond-scale execution while preserving system stability. To support this workflow, \name provides three capabilities:

\begin{enumerate}

\item \noindent \textbf{An autonomous synthesis loop:} This component functions as the system's agentic AI layer, moving beyond static logic to a continuous cycle of hypothesis and experimentation. By treating the live network as the evaluation environment, \name autonomously generates, tests, and evaluates candidate algorithms. It learns from live traces and hardware feedback, allowing the AI agent to refine its logic based on the observed consequences of its decisions.

\item \noindent\textbf{High-fidelity wireless telemetry:} To enable meaningful learning, \name must see what the hardware is actually doing. This component provides the ground-truth visibility required to close the loop between algorithms (code) and performance. It exposes real-time transmission outcomes and underlying hardware behaviors that are typically invisible to the operating system. Without this high-fidelity window into the physical-layer performance, the synthesis loop would remain blind to the hardware idiosyncrasies that dictate real-world success.

\item \noindent\textbf{A programmable environment:} This component provides the agility needed to turn theoretical insights into operational code. By replacing rigid rate controllers with a policy-driven engine, \name can deploy dynamically generated algorithms into the transmission path in under one second. This environment is designed for seamless adaptation, allowing the system to update its logic without disrupting ongoing traffic or disconnecting stations, thereby enabling continuous optimization in a live network.

\end{enumerate}

This paper makes the following contributions:
\begin{enumerate}
    \item \textbf{Autonomous research orchestrator:} We design an LLM-driven agent system capable of independently forming hypotheses, executing live experiments, and interpreting telemetry to refine rate-control logic.
    \item \textbf{Programmable MAC-Layer architecture:} We implement a novel framework that allows the safe, sub-second deployment of machine-generated code into the kernel-space transmission path of commodity Wi-Fi hardware.
    \item \textbf{System validation and case studies:} We demonstrate IteRate’s efficacy by autonomously synthesizing and evaluating multiple rate controllers and conducting automated experiments on a 58-node Wi-Fi testbed. We evaluate \name across \nWorkflows representative workloads capturing interactive voice, web transfers, adaptive video streaming, and bulk downloads. Relative to the well-known and widely used Minstrel algorithm, \name achieves \webImproveMinstrel{} faster web-page loads, \videoImproveMinstrel{} higher video QoE, and \satImproveMinstrel{} higher peak throughput. Compared to policies produced by OpenEvolve, \name reduces web-page completion time by \webImproveOE{} and improves video QoE by \videoImproveOE{}, while increasing peak throughput by \satImproveOE{}. 
    
\end{enumerate}

The software developed in this work will be open-sourced to enable further research.

\section{Background \& Related Work}

\subsection{Challenge of Bitrate Selection}
MAC-layer bitrate selection is the problem of selecting the transmission bitrate (i.e., MCS) used for each wireless frame. Each MCS determines the number of transmitted bits per symbol and the coding rate, which together determine the effective data rate. Higher MCS values transmit more bits per unit time but require a cleaner channel; lower MCS values are more robust but sacrifice throughput and latency.

An ideal rate-maximizing system must operate at the boundary of channel capacity. If the sender overestimates the channel quality and chooses an overly aggressive MCS, frames are corrupted and must be retransmitted, consuming additional airtime and increasing latency. If the sender underestimates the channel and chooses a conservative MCS, transmissions succeed but waste spectrum.

Bitrate adaptation is a control problem under uncertainty: the system must continuously infer the best transmission rate from noisy, delayed feedback while balancing the competing costs of packet loss and underutilized capacity. An ideal system should also adapt to the needs of the applications using the system; e.g., an interactive game might care less about throughput and more about interactive latency (thus favoring bitrates that reduce the number of link-layer retransmissions), while a bulk data transfer might focus on throughput.

\subsection{State-of-the-Art Algorithms}

Existing bitrate algorithms can be categorized into classic heuristics \cite{lacage2004practical,bicket2005bitrate,xia2013minstrel}, reinforcement learning \cite{karmakar2017smartla,karmakar2020banditlink,queiros2022dara}, and machine learning \cite{khastoo2020neura,yin2024adrx}. Most prior work focuses on one of three primary objectives:

\paragraph{1. General-purpose adaptation:} This line of research focuses on developing robust, single-policy solutions capable of handling diverse environments and traffic patterns \cite{lacage2004practical,wong2006robust,xia2013minstrel}. The Linux default, Minstrel \cite{xia2013minstrel}, maintains success statistics for each rate and periodically probes alternatives to identify better options. It selects the bitrate that maximizes the product of the bitrate and success probability at that bitrate. More recent approaches propose RL bandit algorithms such as Thompson Sampling~\cite{karmakar2020banditlink} to accelerate adaptation by balancing exploration and exploitation through systematic uncertainty estimates. These methods rely on deployment-agnostic policies that react to immediate conditions rather than specializing in the specific nuances of a given workload.

\begin{figure}[t]

\begin{subfigure}{\columnwidth}
    \centering
    \includegraphics[width=\textwidth]{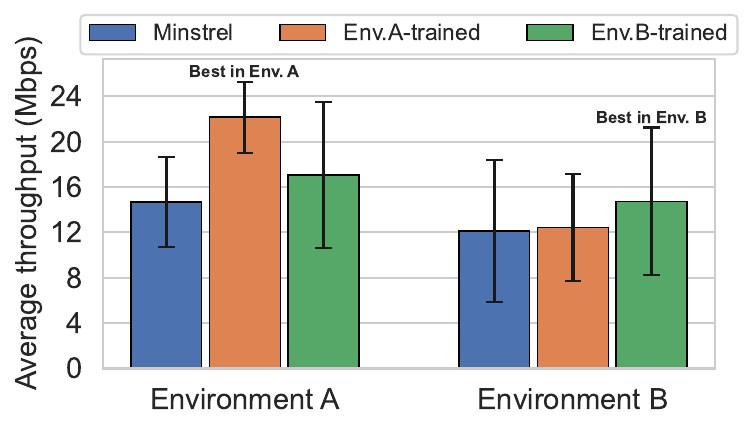}
\end{subfigure}
\caption{\textbf{Bitrate adaptation performance depends on the environment and workload.} \textnormal{Algorithms tuned for one wireless environment often perform poorly in others due to differences in channel conditions and interference.}}
\label{fig:background}

\end{figure}

\paragraph{2. Environment-specific deployment:} A second line of work designs algorithms for a specific deployment setting or hardware regime~\cite{llra,smartrate,raca,damysus,mudra,woof,morpheus,mira,samplelite,ramas,cara,ear,strale,staterate}. For example, LLRA~\cite{llra} targets home networks, SmartRate \cite{smartrate} is built for office/conference-room contention, RaCA \cite{raca} and Damysus \cite{damysus} specialize to dense indoor deployments, and MuDRA \cite{mudra} targets auditorium-scale delivery. Each modifies the bitrate controller to exploit deployment-specific structure that a general-purpose scheme like Minstrel ignores. For instance, in Figure \ref{fig:background}, we present the ability for an algorithm explored in different environments to generalize across environments. In Environment A (a cleaner, more constrained environment), we observe that both of our explored algorithms outperform the default Minstrel algorithm, but the algorithm tuned in that environment performed better. However, in Environment B, only the algorithm that was finetuned was able to substantially outperform Minstrel. 

\paragraph{3. Application-specific:}

The third line of research aims at providing algorithms designed to optimize metrics for specific applications~\cite{yin2024adrx,llra,woof,haratcherev2005streaming,yang2006appaware,choudhury2007multimedia,braskich2005vowlan,choi2008voipra,lee2014vowlan}, in the sense that the ``best'' bitrate policy depends on an application-level objective. Prior work has shown that application performance under a given bitrate scheme varies substantially across workloads such as multimedia streaming, bulk transfer, and web browsing, and that raw link-layer throughput is often an insufficient proxy for end-to-end utility~\cite{haratcherev2005streaming,yang2006appaware}. As a result, some algorithms are designed primarily around packet success probability or burst-loss reduction, others around throughput or expected transmission time, and still others around delay-sensitive or interactive workloads where tail latency, jitter, or conversational quality matter more than raw goodput~\cite{choudhury2007multimedia,llra,braskich2005vowlan,choi2008voipra,lee2014vowlan}. For example, ADR-X aims to reduce packet losses for gaming devices~\cite{yin2024adrx}. Other application-specific or traffic-aware designs target multimedia streaming, VoIP/VoWLAN QoS, and congested mixed-traffic settings~\cite{haratcherev2005streaming,choudhury2007multimedia,braskich2005vowlan,choi2008voipra,lee2014vowlan,woof}. To optimize application-level metrics, these schemes exploit structure in the underlying workload—e.g., deadlines, loss sensitivity, interactivity, or traffic mix—rather than optimizing a single application-agnostic metric~\cite{yang2006appaware,lee2014vowlan,yin2024adrx}. 

Our goal in this paper is to design a system that can autonomously specialize algorithms to the specific deployment environments and application regimes. As we'll show in \S\ref{sec:evaluation}, our system is able to select different algorithms based on how they perform in a unique environment. 

\subsection{Human-in-the-Loop Rate Tuning}

Many Wi-Fi deployments already assume a human-in-the-loop tuning workflow. In these systems, network operators do not treat bitrate behavior as fixed; instead, they inspect telemetry and revise rate-related policies as conditions change. For instance, Extreme's WiNG allows administrators to select specific algorithms \cite{extreme-wing-rate-selection}, while Juniper's Mist provides predefined bitrate profiles and custom settings \cite{mist-data-rates}.

This decision-making process follows a standardized operational pattern: access points (APs) collect telemetry at the edge and export it to a centralized management plane such as Meraki Health \cite{meraki-health}, Mist Cloud \cite{mist-ap-overview}, or Cisco Catalyst Center \cite{cisco-ai-rrm}, where engineers diagnose pathologies and update network configurations.

Recent products from Juniper \cite{mist-rrm}, Cisco \cite{cisco-ai-rrm,meraki-ai-rrm}, and HPE Aruba \cite{aruba-airmatch,aruba-airmatch-brief} automate parts of this loop, using AI/ML to tune radio parameters like channel selection and transmit power. Our work explores whether this same architecture of trace collection and centralized analysis can be pushed much further to the automatic near-real-time synthesis and deployment of MAC-layer bitrate selection algorithms.

\subsection{Code Evolution Challenges}

One natural approach to solve this problem is to use LLM-driven code-mutation (e.g., OpenEvolve \cite{openevolve}) or reasoning pipelines~\cite{glia}. However, existing evolutionary pipelines \cite{openevolve,glia,ADRS} rely on simulators to ensure scoring consistency. While simulators provide the controlled environments necessary to isolate algorithmic performance, they often fail to capture the disorderly, non-stationary realities of physical RF channels and the idiosyncrasies of Wi-Fi hardware~\cite{abedi2014trate,abedi2016tsimn}. Moreover, simulators struggle to faithfully model the causal relationships between low-level bitrate adaptations and behaviors at higher protocol layers because small changes can propagate into complex network-level effects \cite{alomar2023causalsim}.

Our approach is instead to evolve the bitrate selection algorithm directly \emph{online} in the actual wireless environment. However, the key technical challenge is that prior works, such as OpenEvolve, rely solely on end-to-end scoring functions to grade candidate programs. This approach works well in controlled, reproducible simulations. However, in real wireless environments, changing traffic patterns and transient interference introduce environmental noise, causing such evolutionary search to degenerate into a near-random walk.

\begin{figure}
    \centering    \includegraphics[width=\columnwidth]{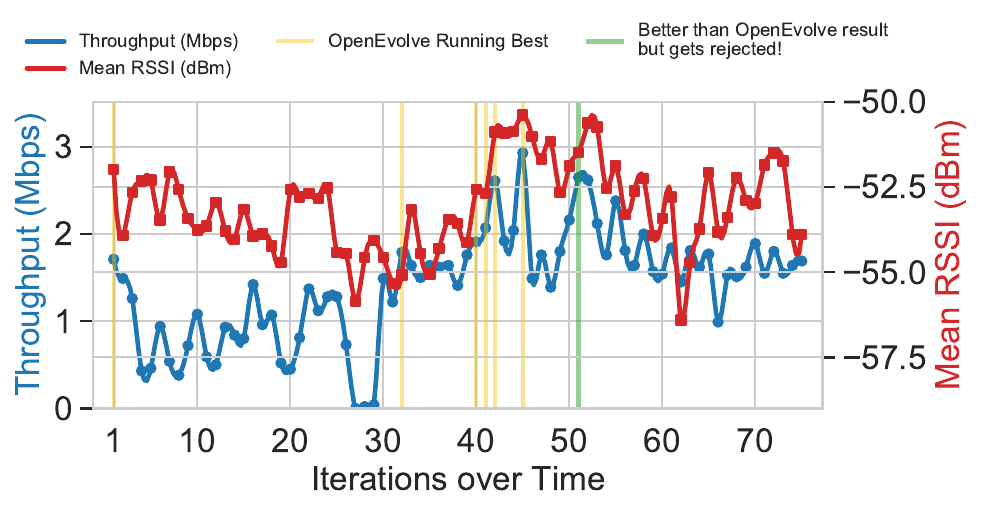}
    \caption{Environmental noise corrupts evolutionary scoring in live wireless settings. \textnormal{Throughput (blue) closely tracks RSSI (red) across OpenEvolve iterations, causing protocols evaluated during favorable periods to appear superior. OpenEvolve selects protocols due to RSSI improvements (yellow lines) rather than protocol gains. A random protocol evaluated under worse channel conditions (green marker) achieves higher true performance but is rejected, illustrating how channel variability can mislead evolutionary search.}}
    \label{fig:openevolve}
\end{figure}

Fig.~\ref{fig:openevolve} illustrates this effect using an OpenEvolve run where each iteration evaluates a candidate protocol on a live wireless link (\S\ref{sec:implementation}) while recording both the application metric (throughput, blue) and channel quality (RSSI, red). Because throughput closely tracks RSSI, protocols evaluated during favorable channel periods appear superior even when the improvement is unrelated. OpenEvolve selects candidates based on transient channel conditions rather than genuine algorithmic gains. In fact, when we later tested a randomly generated protocol, we observed that it could outperform the previously selected ``best'' protocol. 

Instead, as we show in the next section, our goal is to equip the evolutionary agent with an explicit understanding of the wireless environment so it can reason about \emph{why a method works, not just whether it performs well}. By embedding domain insights similar to a network researcher, the agent can make more principled design decisions, adapt more robustly to changing conditions, and avoid discarding good strategies due to transient network fluctuations. This is the same philosophy advocated in Glia~\cite{glia}, but the differences lie in the agentic structure and in the use of the live network as the experimentation platform rather than a simulator.

\section{\name Architecture}
\label{sec:system}

\begin{figure}[t]
    \centering
    \resizebox{\columnwidth}{!}{\begin{tikzpicture}[
    >=stealth,
    font=\sffamily,
    box/.style={draw=none, rounded corners=8pt, fill=#1, text centered},
    server/.style={box=blue!10, minimum width=6.2cm, minimum height=1.8cm},
    broker/.style={box=orange!15, minimum width=4.4cm, minimum height=0.8cm},
    device/.style={box=teal!15, minimum width=2.0cm, minimum height=2.2cm},
    innerbox/.style={draw=none, fill=white, rounded corners=4pt, font=\sffamily\scriptsize, minimum width=1.6cm, minimum height=0.45cm, text centered},
    serverinner/.style={innerbox, minimum width=2.6cm},
    lbl/.style={font=\sffamily\scriptsize\bfseries, text=black!70, fill=white, inner sep=2pt},
]

\node[server] (server) at (0, 0) {};
\node[font=\sffamily\small\bfseries, text=black!85, anchor=north] at ([yshift=-4pt]server.north) {Agent Server};
\node[serverinner, text=black!75] (llm) at (-1.4, -0.4) {LLM Orchestrator};
\node[serverinner, text=black!75] (store) at (1.4, -0.4) {Session Store};

\node[broker, font=\sffamily\small\bfseries, text=black!85] (broker) at (0, -1.9) {MQTT Broker};

\node[font=\large, text=black!20] at (-4.0, -4.2) {$\cdots$};

\node[device] (r1) at (-2.5, -4.2) {};
\node[font=\sffamily\scriptsize\bfseries, text=black!85, anchor=north] at ([yshift=-4pt]r1.north) {Wi-Fi Device};
\node[innerbox, text=black!75] (k1) at (-2.5, -4.0) {Kernel Module};
\node[innerbox, text=black!75] (d1) at (-2.5, -4.6) {Mgmt Daemon};

\node[device] (r2) at (0, -4.2) {};
\node[font=\sffamily\scriptsize\bfseries, text=black!85, anchor=north] at ([yshift=-4pt]r2.north) {Wi-Fi Device};
\node[innerbox, text=black!75] (k2) at (0, -4.0) {Kernel Module};
\node[innerbox, text=black!75] (d2) at (0, -4.6) {Mgmt Daemon};

\node[device] (r3) at (2.5, -4.2) {};
\node[font=\sffamily\scriptsize\bfseries, text=black!85, anchor=north] at ([yshift=-4pt]r3.north) {Wi-Fi Device};
\node[innerbox, text=black!75] (k3) at (2.5, -4.0) {Kernel Module};
\node[innerbox, text=black!75] (d3) at (2.5, -4.6) {Mgmt Daemon};

\node[font=\large, text=black!20] at (4.0, -4.2) {$\cdots$};

\draw[->, line width=1.2pt, cyan!80!blue] (-0.5, -0.9) -- (-0.5, -1.5)
    node[midway, left, lbl] {eBPF ELF, cmds};
\draw[->, line width=1.2pt, magenta!80!red] (0.5, -1.5) -- (0.5, -0.9)
    node[midway, right, lbl] {telemetry, stats};

\draw[->, line width=1.2pt, cyan!80!blue] (-1.1, -2.3) -- (-2.1, -3.1);
\draw[->, line width=1.2pt, magenta!80!red] (-1.8, -3.1) -- (-0.8, -2.3);

\draw[->, line width=1.2pt, cyan!80!blue] (-0.15, -2.3) -- (-0.15, -3.1);
\draw[->, line width=1.2pt, magenta!80!red] (0.15, -3.1) -- (0.15, -2.3);

\draw[->, line width=1.2pt, cyan!80!blue] (1.1, -2.3) -- (2.1, -3.1);
\draw[->, line width=1.2pt, magenta!80!red] (1.8, -3.1) -- (0.8, -2.3);

\end{tikzpicture}}
    \caption{System topology. \textnormal{Wi-Fi devices execute rate policies and stream per-frame telemetry to the centralized agent server, which hosts the LLM orchestrator that synthesizes algorithms and deploys compiled eBPF objects to the edge.}}
    \label{fig:system-diagram}
\end{figure}
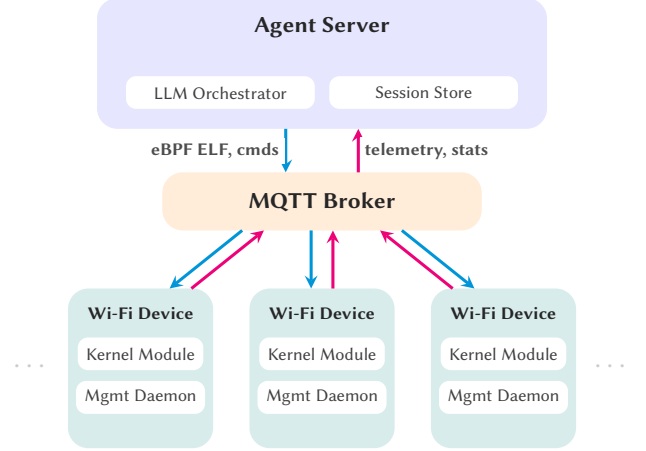

\name is an autonomous, LLM-driven research platform that dynamically synthesizes, deploys, and evaluates MAC-layer rate-adaptation algorithms directly on a wireless network. As shown in Figure~\ref{fig:system-diagram}, the system consists of a centralized management server and a fleet of distributed Wi-Fi devices. The devices act as execution engines that run the selected bitrate policies and stream telemetry back to the server. All computationally intensive tasks (i.e. the agent discovering and refining new algorithms) are handled centrally by an autonomous research orchestrator.

\noindent\textbf{Safety.} The architecture enforces strict isolation to enable safe experimentation without compromising stability. \name separates the compute-intensive, non-deterministic reasoning of the LLM from the critical on-device software executing the policy in real-time. This separation is essential because LLMs may hallucinate or make logical errors, and even minor hardware misconfigurations can trigger network outages that take minutes to recover from. To safely deploy an autonomous agent in low-level network control, \name implements a three-layer safety model:
\begin{enumerate}
\item All agent-generated code must pass the kernel’s eBPF static verifier before execution, ensuring memory safety, bounded runtime, and protection against kernel crashes.
\item The reasoning agent cannot directly modify device state, ensuring strict privilege separation.
\item All configuration changes are recorded in an ephemeral undo log and automatically reverted on shutdown, preventing persistent misconfiguration (\S\ref{sec:implementation}). 
\end{enumerate}

\subsection{The Autonomous Research Agent}

\name's design mimics the workflow and intuition of wireless networking researchers. The \name agent does not optimize metrics blindly; instead, it observes link behavior, forms hypotheses about channel dynamics and protocol interactions, designs algorithmic modifications grounded in those observations, and tests them against empirical evidence. 

Like a human expert, the agent reasons about causality, iterates on promising ideas, and abandons ineffective ones based on principled analysis rather than noisy outcomes. The goal is to embed the investigative, hypothesis-driven skill set of wireless networking researchers into an automated, scalable system.

\begin{figure*}[t]
    \centering
    \resizebox{\textwidth}{!}{\begin{tikzpicture}[
    >=stealth,
    font=\sffamily,
    basebox/.style={draw=none, rounded corners=6pt, text centered},
    scientist/.style={basebox, fill=blue!10, minimum width=2.8cm, minimum height=3.6cm},
    subagent/.style={basebox, fill=green!10, rounded corners=4pt, minimum width=3.2cm, minimum height=0.6cm, font=\sffamily\scriptsize},
    infra/.style={basebox, fill=orange!15, minimum width=2.0cm},
    store/.style={basebox, fill=yellow!15, rounded corners=4pt, minimum width=3.4cm, minimum height=0.6cm, font=\sffamily\scriptsize},
    edgecontainer/.style={basebox, fill=teal!15, rounded corners=6pt},
    lbl/.style={font=\sffamily\scriptsize\bfseries, text=black!70, fill=white, inner sep=2pt, rounded corners=3pt},
    privilegelbl/.style={lbl, text=red!50, draw=red!20, line width=0.8pt},
]

\node[scientist] (sup) at (0, -0.3) {};
\node[font=\sffamily\small\bfseries, text=black!85, anchor=north] at (0, 1.3) {Scientist};

\begin{scope}[every node/.style={font=\sffamily\scriptsize, text=black!75, anchor=west}]
    \node at (-0.9, 0.6)   {1. Orient};
    \node at (-0.9, 0.2)   {2. Hypothesize};
    \node at (-0.9, -0.2)  {3. Design};
    \node at (-0.9, -0.6)  {4. Execute};
    \node at (-0.9, -1.0)  {5. Interpret};
    \node at (-0.9, -1.4)  {6. Iterate};
\end{scope}

\draw[densely dashed, gray, line width=1.2pt, opacity=0.5] (2.3, 1.8) -- (2.3, -3.4);
\node[privilegelbl, draw=none, text=gray, anchor=south] at (2.3, 1.8) {privilege boundary};

\node[subagent] (sa1) at (4.8, 0.9) {Experiment Runner};
\node[subagent] (sa2) at (4.8, 0.1) {Algorithm Designer};
\node[subagent] (sa3) at (4.8, -0.7) {Data Analyst};
\node[subagent] (sa4) at (4.8, -1.5) {Network Engineer};

\draw[->, line width=1.2pt, cyan!80!blue] ([yshift=6pt]sup.east |- sa1) -- ([yshift=6pt]sa1.west) node[pos=0.23, above=0pt, lbl, fill=none] {\texttt{task()}};
\draw[<-, line width=1.2pt, magenta!80!red, densely dashed] ([yshift=-6pt]sup.east |- sa1) -- ([yshift=-6pt]sa1.west) node[pos=0.78, below=0pt, lbl, fill=none] {results};

\draw[->, line width=1.2pt, cyan!80!blue] (sup.east |- sa2) -- (sa2.west);
\draw[->, line width=1.2pt, cyan!80!blue] (sup.east |- sa3) -- (sa3.west);
\draw[->, line width=1.2pt, cyan!80!blue] (sup.east |- sa4) -- (sa4.west);

\node[infra, minimum height=2.8cm] (tools) at (8.0, -0.3) {};
\node[font=\sffamily\small\bfseries, text=black!85, anchor=north] at (8.0, 0.9) {Tool API};
\node[font=\sffamily\scriptsize, text=black!70] at (8.0, 0.35) {90+ tools};
\node[font=\sffamily\scriptsize, text=black!70] at (8.0, 0.10) {15 groups};
\node[font=\sffamily\scriptsize, text=black!50] at (8.0, -0.5) {\textit{observe}};
\node[font=\sffamily\scriptsize, text=black!50] at (8.0, -0.75) {\textit{actuate}};
\node[font=\sffamily\scriptsize, text=black!50] at (8.0, -1.0) {\textit{stimulate}};

\draw[->, line width=1.2pt, cyan!80!blue] (sa1.east) -- (tools.west |- sa1);
\draw[->, line width=1.2pt, cyan!80!blue] (sa2.east) -- (tools.west |- sa2);
\draw[->, line width=1.2pt, cyan!80!blue] (sa3.east) -- (tools.west |- sa3);
\draw[->, line width=1.2pt, cyan!80!blue] (sa4.east) -- (tools.west |- sa4);

\node[infra, minimum height=0.7cm, minimum width=2.6cm, font=\sffamily\small\bfseries, text=black!85] (mqtt) at (11.2, 0.5) {MQTT Broker};

\node[edgecontainer, minimum height=1.9cm, minimum width=2.6cm] (edge) at (11.2, -1.3) {};
\node[font=\sffamily\small\bfseries, text=black!85, anchor=north] at (11.2, -0.45) {Wi-Fi Devices};
\node[yshift=-5pt, draw=none, fill=white, rounded corners=2pt, font=\sffamily\scriptsize, text=black!70, minimum width=2.3cm, minimum height=0.35cm] at (11.2, -1.7) {Kernel Module};
\node[yshift=-5pt, draw=none, fill=white, rounded corners=2pt, font=\sffamily\scriptsize, text=black!70, minimum width=2.3cm, minimum height=0.35cm] at (11.2, -1.1) {Mgmt Daemon};

\draw[->, line width=1.2pt, cyan!80!blue] (tools.east |- mqtt) -- (mqtt.west);
\draw[<-, line width=1.2pt, magenta!80!red] (tools.east |- edge) -- (edge.west);

\draw[->, line width=1.2pt, cyan!80!blue] ([xshift=-15pt]mqtt.south) -- ([xshift=-15pt]edge.north) node[midway, left=3pt, lbl] {cmds};
\draw[<-, line width=1.2pt, magenta!80!red] ([xshift=15pt]mqtt.south) -- ([xshift=15pt]edge.north) node[midway, right=3pt, lbl] {telem};

\node[store] (db) at (4.8, -3.0) {Session Store (SQLite)};

\draw[->, line width=1.2pt, magenta!80!red, rounded corners=4pt] ([xshift=15pt]sup.south) |- ([yshift=5pt]db.west) node[pos=0.65, above, lbl, fill=none] {findings};

\draw[->, line width=1.2pt, cyan!80!blue, densely dashed, rounded corners=4pt] ([yshift=-6pt]db.west) -| ([xshift=-15pt]sup.south) node[pos=0.25, below, lbl] {\textit{query}};

\draw[->, line width=1.2pt, magenta!80!red, rounded corners=4pt] (tools.south) |- (db.east) node[pos=0.75, above, lbl] {telemetry};

\end{tikzpicture}}
    \caption{Agentic pipeline. \textnormal{The Scientist agent drives the research loop but cannot mutate device state. It delegates operational tasks across a privilege boundary to scoped subagents, each with access to a restricted subset of the 90+ tool API. Raw telemetry flows directly into the session store, bypassing the LLM context.}}
    \label{fig:agentic-diagram}
\end{figure*}
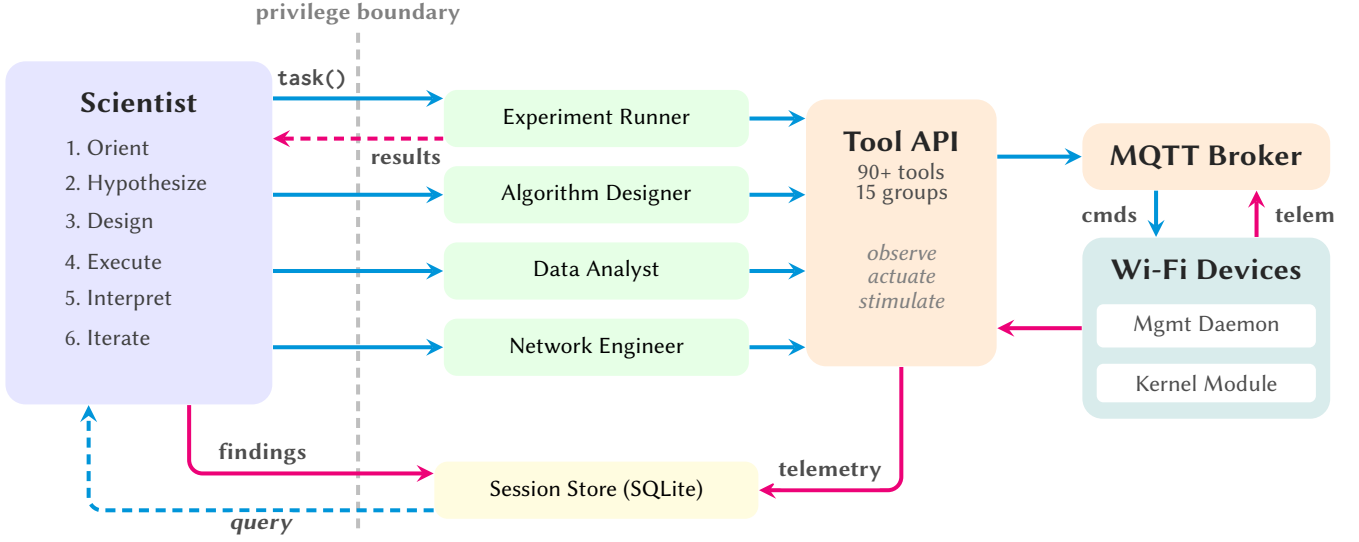

At the top of our system is a unified \emph{Scientist} tool calling agent responsible for the full research loop. It designs algorithms, plans experiments, deploys them, collects data, and analyzes results.

\subsubsection{Scientist State Machine}
To ensure this exploration process remains structured and reliable, we model the research workflow as a growing process. Since unbounded LLM agents are prone to logical drift and stalled execution \cite{agent_drift}, the Scientist agent is prompted to follow a flexible six-step research protocol that formalizes the progression from hypothesis to validated results:

\begin{enumerate}[leftmargin=*, labelindent=0pt]
\item \textbf{Orient:} Query testbed status, review prior findings from session memory, and search external academic literature.
\item \textbf{Hypothesize:} Propose a formally testable hypothesis containing a falsifiable statement, a specific performance prediction, and a physical rationale.
\item \textbf{Design:} Formulate an experimental plan isolating independent and dependent variables (e.g., isolating payload size while varying the MCS).
\item \textbf{Execute:} Delegate the operational plan to an Experiment Runner subagent (described below), which orchestrates simultaneous traffic generation and background telemetry collection.
\item \textbf{Interpret:} A Data Analyst subagent (described below) computes effective metrics; the Scientist writes a formalized finding supported by specific empirical evidence.
\item \textbf{Iterate:} Update the hypothesis status in the database and refine the approach based on accumulated evidence.
\end{enumerate}

To execute each step, the Scientist invokes a dedicated tool that records the action, intermediate reasoning, and conclusions to a persistent store (\S\ref{sec:system_memory}).

\subsubsection{Hierarchical Subagent Topology}

To execute each of these steps, the Scientist requires different expertise, and each step entails different risks, especially because experiments are conducted on live networks. To manage these tasks, we decompose the Scientist into structured subagents, each handling a specific function. This separation reduces risk, clarifies responsibility, and manages the Scientist's context window while keeping the overall process coordinated as a single research workflow:

\begin{itemize}[leftmargin=*, labelindent=0pt]
\item \textbf{Experiment Runner:} Orchestrates the physical testbed by executing A/B tests, commanding throughput sweeps, and managing background telemetry collection.
\item \textbf{Algorithm Designer:} A synthesis agent that operates offline to produce ideas and express them in eBPF C code, adhering to kernel verifier constraints.
\item \textbf{Data Analyst:} Computes effective metrics, identifies performance anomalies (e.g., MCS knee points), and generates cross-environment comparisons from stored telemetry.
\item \textbf{Network Engineer:} Manages physical and logical hardware configurations via configuration updates that are automatically reverted on session teardown.
\end{itemize}

Each subagent is kept as a stateless agent with a unique, fixed tool subset (breakdowns in Appendix \ref{app:agent}). The Scientist delegates via a \texttt{task(description, subagent\_type)} tool call, passing a natural-language task description that includes all required context. Because subagents are stateless (fresh LLM context per invocation), the Scientist must include all relevant context in each call, while subagents return structured JSON results. This design isolates strategic reasoning from operational execution and ensures that no subagent accumulates stale state across tasks.

\subsubsection{Tool-Use API and Data Delegation}
All subagents interface with the physical network via a tool server that exposes over 90 dynamically loadable tools, organized into 15 capability groups (rate control, telemetry, traffic generation, eBPF management, experiment orchestration, etc.). Tools map high-level intent into deterministic MQTT commands across three vectors: \emph{observation} (telemetry querying and statistics), \emph{actuation} (rate configuration, policy switching, BPF deployment), and \emph{stimulus} (coordinated bidirectional traffic generation).

A critical design constraint is that MAC-layer telemetry generates tens of thousands of per-frame records per minute, which would quickly exhaust any current LLM's context window. \name enforces strict data delegation: raw telemetry payloads never pass through the LLM context. Instead, telemetry is streamed directly from edge devices to a per-session relational datastore. The analyst agent assesses this data through tool calls that return compact statistical summaries.

\subsubsection{Relational Memory and Cross-Session Lineage} \label{sec:system_memory}

To support long-horizon research beyond the LLM’s limited context window, the Scientist is equipped with a tool that commits intermediate results and learned insights to a persistent SQLite datastore. The agent is instructed to record its findings after each step of the research loop, creating a structured, relational memory that survives across sessions and enables cumulative progress.

\subsubsection{Execution} During execution, the agent generates eBPF code that complies with kernel verifier constraints (e.g., bounded loops, controlled stack usage, pinned maps). The code is cross-compiled centrally and deployed to the router (\S\ref{sec:pipeline}). The kernel verifier then validates the program before it is loaded. If validation fails, the verifier’s error trace is returned to the algorithm designer subagent for automated debugging. This execution pipeline ensures that only safe, verifiable programs run on the device while enabling rapid iteration.

\section{Implementation}
\label{sec:implementation}

\name runs bitrate algorithms directly inside the Linux Wi-Fi transmission datapath, allowing the system to observe link outcomes and update transmission policies at frame-level granularity. Traditionally, Linux Wi-Fi drivers rely on a pluggable \emph{rate control module} that selects the modulation and coding scheme (MCS). We replace the default Minstrel-HT controller with a programmable module that allows the centralized agent to deploy and update rate-selection algorithms on a 58-node commodity hardware testbed.

The bitrate algorithm periodically runs and updates a stored MCS lookup table. Then, before a frame is transmitted, the kernel consults this table to select the transmission rate. After the transmission completes, the driver reports the outcome (success or failure, retry count, and signal strength). Our module exposes this per-frame feedback to a policy engine, which updates its internal state and determines the rate for subsequent transmissions. Policies may be implemented as built-in algorithms or as dynamically loaded eBPF programs, enabling the agent to deploy new rate-selection logic without rebooting devices or modifying firmware.

The implementation consists of five components: 
(1) a kernel-level programmable rate control module, 
(2) an in-kernel eBPF execution environment for policy logic, 
(3) a telemetry subsystem for collecting transmission outcomes, 
(4) a lightweight management daemon that bridges devices to the centralized agent, and 
(5) an automated build and deployment pipeline for synthesizing and deploying new algorithms across the testbed. 

Together, these components enable the agent to observe link behavior, generate candidate algorithms, deploy them safely to edge devices, and evaluate their performance in real wireless environments.

\subsection{Programmable Rate Control Module}
\label{sec:rc-module}

The edge datapath in \name is a custom Linux \texttt{mac80211} rate control module that replaces the default Minstrel-HT algorithm. The module registers as a standard \texttt{rate\_control\_ops} provider and implements the \texttt{get\_rate()} and \texttt{tx\_status()} callbacks that the Wi-Fi stack invokes before each frame transmission and after each transmission completion, respectively. This placement allows the system to observe per-frame transmission outcomes and update rate decisions at frame-level granularity.

\noindent\textbf{Policy Engine.} Rather than enforcing a single fixed algorithm, the module implements a \emph{policy dispatch} architecture with five interchangeable policies:

\begin{itemize}[leftmargin=*, labelindent=0pt, itemsep=1pt]
\item \textbf{Fixed:} A single MCS for all stations and frame types.
\item \textbf{Per-station:} Individual rate overrides per MAC address, enabling controlled per-link experiments.
\item \textbf{Round-robin:} Cycles through a configurable MCS list, transmitting $N$ frames at each rate before advancing. This policy is used for systematic rate sweeps and training data collection.
\item \textbf{Frame-type:} Separate rates for management, control, and data frames, allowing the agent to protect control-plane reliability while aggressively tuning data rates.
\item \textbf{BPF:} Rates driven by an eBPF array map, enabling fully autonomous in-kernel rate adaptation (\S\ref{sec:bpf-env}).
\end{itemize}

Policies can be switched at runtime via a debugfs control interface without disrupting active associations. To avoid pushing rate tables on every frame, the module maintains a global generation counter. When the configuration changes, the generation counter increments, and the per-frame hot path updates the driver's rate table only when a station's cached generation becomes stale, amortizing update costs across thousands of frames.

\noindent\textbf{Rate Table Integration.} Wi-Fi drivers typically support two mechanisms for supplying transmission rates: a \emph{per-packet path}, where the rate is embedded in each frame's TX descriptor, and a \emph{cached rate table path}, where rates are pushed once into the driver's per-station cache and reused for subsequent frames. On our hardware, we observe that the per-packet path produces unreliable behavior at higher MCS indices. We therefore use the cached rate table mode, ensuring consistent rate enforcement across transmissions.


\subsection{eBPF Execution Environment}
\label{sec:bpf-env}

Rate adaptation decisions occur on microsecond timescales within the transmission path, making user-space control too slow due to scheduling latency and kernel-user round trips. To support real-time adaptation, the BPF policy mode executes rate-selection logic directly inside the kernel's TX completion path. On each frame completion, the program observes transmission outcomes and updates rate decisions at frame granularity.

Before deployment, each eBPF program is statically verified by the Linux kernel to guarantee memory safety and bounded execution. The verifier ensures that programs terminate, access only valid memory regions, and respect stack limits, preventing faulty policies from destabilizing the device. This allows dynamically generated algorithms to execute safely in the datapath.

\noindent\textbf{Shared Maps.} The module exposes three persistent BPF array maps, each pinned to \texttt{/sys/fs/bpf/} for cross-process sharing between the kernel datapath and the management daemon:

\begin{itemize}[leftmargin=*, labelindent=0pt, itemsep=1pt]

\item \textbf{Rate Map} (128 entries $\times$ 8\,B): Indexed by wireless client ID (WCID). Each entry specifies the target transmission parameters (MCS, spatial streams, bandwidth, guard interval, PHY mode) and a valid bit. The kernel reads this map on every \texttt{get\_rate()} invocation to determine the next transmission rate. Because BPF map writes do not increment the module's \texttt{rate\_generation} counter, the BPF policy handler independently detects rate changes by comparing the current map entry against the previously applied rate and resetting the push state when they differ.

\item \textbf{Stats Map} (128 entries $\times$ 48\,B): The kernel periodically updates this map with aggregated transmission statistics, batching writes every 64 TX completions per station. Entries include cumulative transmission counts, an EWMA packet error rate (PER), retry counts, and signal strength. These summaries provide BPF programs with statistical context while minimizing datapath overhead.

\item \textbf{Algorithm Map} (128 entries, variable size): Private per-station state used by rate adaptation algorithms (e.g., per-MCS delivery histograms, probing counters, or EWMA windows). This map is created and pinned by the management daemon using raw \texttt{bpf()} syscalls to avoid BTF type dependencies absent on our embedded MIPS kernels.

\end{itemize}

Map \emph{pointer} swaps (e.g., when the management daemon attaches a new map or detaches an old one) are RCU-protected. The module publishes new map references using \texttt{rcu\_assign\_pointer()} and waits for a grace period before releasing old maps, allowing the per-frame TX path to read map pointers lock-free via \texttt{rcu\_dereference()}. Individual map entry updates rely on the BPF subsystem's atomic operations and therefore do not require RCU.

\noindent\textbf{Per-Frame TX Context.} When a BPF program is attached, the kernel invokes it after each frame completion with a 120-byte context structure containing 15 fields (Table~\ref{tab:bpf-ctx}). The context includes transmission outcomes, retry counts, signal strength, and both \emph{configured} and \emph{hardware-reported} rate information.

The \texttt{hw\_mcs\_used} and \texttt{hw\_rate\_flags} fields expose the actual rate used by the radio firmware rather than the rate requested by the module. This distinction is important because the MT76x02 radio automatically falls back to lower rates when retries are exhausted, including cross-PHY transitions from high throughput (HT) to legacy OFDM. The standard mac80211 rate control interface does not expose this fallback rate separately. Our module captures the hardware-reported rate directly from the driver's TX status before any subsequent processing. Without this visibility, rate adaptation algorithms attribute successful delivery to the configured MCS even when the frame was recovered by firmware fallback, grossly overestimating the reliability of aggressive rates.

\noindent\textbf{Verifier Safety.} All eBPF programs must pass the kernel's static verifier before execution. The verifier guarantees bounded execution time (no unbounded loops), memory safety (no out-of-bounds accesses), and compliance with the BPF stack limit ($\leq$512 bytes).

A subtle constraint arises on embedded kernels without BPF Type Format (BTF) support: values loaded from BPF maps are treated by the verifier as unconstrained scalars. For example, a \texttt{\_\_u8} MCS index loaded from the algorithm map and used as an array index is conservatively treated as a value in the range (0,\,255), even if the program logic guarantees a smaller bound. Developers must therefore emit explicit bounds checks before every map-derived array access. We encode this constraint in the static linter described in \S\ref{sec:pipeline}.

\subsection{Telemetry Subsystem}
\label{sec:telemetry}

High-fidelity telemetry is essential for closing the agent's observe-hypothesize-experiment loop. The module maintains a per-PHY lock-protected ring buffer of 4,096 entries (approximately 80\,KB), recording a compact 20-byte entry for every transmitted frame:

\begin{itemize}[leftmargin=*, labelindent=0pt, itemsep=1pt]
\item \textbf{Intended rate:} The MCS index and flags configured by the active policy.
\item \textbf{Hardware rate:} The actual rate reported by the driver in TX status. The module captures this from the raw status structure before any subsequent processing.
\item \textbf{Outcome:} Success/failure, retry count, aggregation flag.
\item \textbf{Metadata:} WCID, frame length, RSSI, monotonic sequence number, kernel timestamp.
\end{itemize}

The ring buffer is read by the management daemon through a binary debugfs interface. Each read atomically snapshots the head and tail pointers, copies available entries (handling ring wraparound with two \texttt{memcpy} calls), and advances the tail, providing zero-copy transfer to userspace. At typical traffic rates (1,000--10,000 frames/s), the buffer provides 0.4--4\,s of history before wraparound. The management daemon polls at 1\,s intervals; at the high end of this range, some entries may be overwritten before they are read. This is acceptable because the BPF program's algorithm map maintains its own cumulative per-station statistics independently of the telemetry ring, and the agent's analytical tools operate on statistical aggregates rather than individual frame records. 


\subsection{Management Daemon}
\label{sec:daemon}

Each Wi-Fi device runs a lightweight management daemon ($\approx$8k lines of C) that bridges the kernel module to the centralized agent over Message Queuing Telemetry Transport (MQTT). The daemon operates as a single-threaded asynchronous event loop.

\noindent\textbf{Command Routing.} The daemon subscribes to over 40 MQTT command topics covering rate configuration, BPF map management, wireless interface lifecycle, traffic generation, packet capture, and Unified Configuration Interface (UCI). Each command produces a structured acknowledgment containing the result or a detailed error message, enabling the agent to confirm that changes were applied before proceeding with experiments.

\noindent\textbf{BPF Lifecycle.} The daemon manages BPF artifacts using a dual-library strategy dictated by embedded kernel constraints. Map creation uses raw \texttt{bpf()} syscalls to avoid BTF type dependencies absent on MIPS targets; program loading uses libbpf for ELF parsing, relocation, and map binding. Programs arrive as base64-encoded ELF objects over MQTT and are loaded asynchronously on a background thread. On verifier failure, libbpf's output callback captures the verifier log (truncated to 3\,KB) and returns it to the agent for autonomous debugging.

\noindent\textbf{Telemetry Streaming.} On a configurable interval (default 1\,s), the daemon reads the kernel's binary telemetry ring buffer, converts entries to JSON, and publishes them to the MQTT broker. A separate stats publication (default 5\,s) provides per-station aggregate statistics including delivery ratios, retry distributions, and signal strength. Raw telemetry flows directly into the agent's relational datastore without passing through the LLM context window.

\noindent\textbf{Ephemeral Configuration.} All UCI configuration changes (wireless interface creation, network settings) are tracked in an in-memory undo log. On daemon shutdown or session teardown, all changes are automatically reverted, preventing permanent device misconfiguration from failed experiments. Individual changes can be selectively persisted or reverted via MQTT commands.

\subsection{Build and Deployment Pipeline}
\label{sec:pipeline}

When the agent synthesizes a new eBPF rate control algorithm, the following pipeline executes autonomously:

\begin{enumerate}[leftmargin=*, labelindent=0pt, itemsep=1pt]
\item \textbf{Synthesis.} The algorithm designer subagent writes eBPF C source, adhering to verifier constraints: compile-time loop unrolling (\texttt{\#pragma unroll}), bounded stack allocation, pinned map references, and explicit bounds checks on all map-derived array indices.
\item \textbf{Static Lint.} A five-rule linter scans the source before compilation, catching the most common causes of verifier rejection:
\emph{(1)}~array accesses indexed by map-loaded values without a preceding bounds check;
\emph{(2)}~loops missing a \texttt{\#pragma unroll};
\emph{(3)}~functions lacking \texttt{\_\_always\_inline} (required on kernels without BPF-to-BPF call support);
\emph{(4)}~excessive conditional nesting on map fields (threshold: 8 branches), which risks verifier state explosion;
\emph{(5)}~stack-allocated structs exceeding the 512-byte BPF stack limit.
The linter returns structured warnings with line numbers; the algorithm designer uses these to correct the source before compilation.
\item \textbf{Cross-Compilation.} The source is compiled targeting BPF EL at optimization level 2, using the target platform's kernel headers to produce a relocatable ELF object.
\item \textbf{OTA Deployment.} The ELF binary is base64-encoded and transmitted over MQTT to the target device. The daemon idempotently initializes BPF maps (if not already pinned), loads the program via libbpf, and attaches it to the kernel module's TX status hook or timer-based invocation path.
\item \textbf{Verification Feedback.} If the kernel verifier rejects the program, the verifier trace is returned to the agent. The algorithm designer diagnoses the failure (typically a missing bounds check or an unrolled loop exceeding the instruction budget) and generates a corrected version.
\item \textbf{Activation.} On successful load, the daemon switches the kernel module to BPF policy. The program begins executing on each TX completion, reading the stats map and writing rate decisions to the rate map.
\end{enumerate}

The end-to-end latency from source code to in-kernel execution is under 10 seconds, dominated by cross-compilation. Source code and compiled ELF binaries are transmitted directly between the build system and edge devices without passing through the LLM context, preserving the agent's context budget for reasoning.

\subsection{Testbed}
\label{sec:testbed}

We deploy \name on a fleet of 58 Hak5 Wi-Fi Pineapple Mark~VII devices distributed across multiple floors of an urban university campus building. Each device is a commodity 802.11n router running OpenWrt with a Linux 6.6 kernel, equipped with a MediaTek MT76x02 2.4\,GHz radio. The devices operate in HT mode with MCS 0--7 on a single spatial stream at 20\,MHz bandwidth, providing PHY rates from 6.5 to 65\,Mbps. Link distances are highly variable, ranging from adjacent rooms on the same floor to cross-floor placements separated by concrete and steel, producing a wide spread of signal conditions. The 2.4\,GHz band is shared with the building's production Wi-Fi infrastructure and neighboring networks, exposing experiments to realistic co-channel interference.

The centralized agent server communicates with all devices over the campus Wi-Fi network via an MQTT broker. Each device runs the custom kernel module and management daemon, with the stock Minstrel-HT rate controller disabled and replaced by our policy-driven module. The \name agent currently uses OpenAI's GPT-5.4 for all LLM inference, accessed via the LangGraph API.

\section{Evaluation}
\label{sec:evaluation}

We evaluate \name's ability to evolve bitrate selection logic in real-time. We compare its performance against state-of-the-art algorithms across diverse workloads and channel conditions, and analyze the computational overhead of the online AI agent.

We allow the agent to converge to an algorithm for a maximum of 50 iterations of the loop. For all results, we report post-convergence A/B tests across 15 paired device samples, rotating between our algorithm and the baselines across multiple links for 2 minutes per sample. 

\textbf{Workloads and QoE Metrics:}
We evaluate the system using \nWorkflows workloads that represent common wireless applications. Traffic is created by synthetically generating traffic patterns from a tool designed to use \texttt{iperf} to model application-layer traffic~\cite{mami_trafic}. This allows us to systematically vary transmission patterns and load characteristics to produce controlled and reproducible traces. Each workload is paired with a distinct application-layer quality-of-experience (QoE) metric, reflecting the performance objective that matters to that application.
\begin{enumerate}[leftmargin=*, labelindent=0pt, itemsep=1pt]

\item \textbf{Peak Throughput.}
A single TCP flow runs without a bandwidth cap for 10\,s, fully loading the wireless link. 
The QoE metric is receiver-side goodput (Mbps), which captures the maximum throughput achievable under sustained load.

\item \textbf{File Download.}
We emulate bulk transfers using TCP bursts of 25\,MB repeated three times. 
The QoE metric is average goodput across transfers, representing the effective download rate experienced by users during large file transfers.

\item \textbf{Web Page Load.}
To approximate web browsing, we generate TCP bursts of 1{,}246\,KB (the median web page size reported by HTTP Archive 2024). 
The QoE metric is mean flow completion time (FCT), which approximates the latency a user experiences when loading a page.

\item \textbf{Voice over IP (VoIP).}
We emulate conversational voice traffic using a UDP stream sending G.711-sized packets (160\,B every 20\,ms) for 30\,s. 
The QoE metric is Mean Opinion Score (MOS), computed from packet loss, jitter, and delay using a simplified ITU-T G.107 E-model.

\item \textbf{Adaptive Video Streaming.}
We model video streaming using repeated TCP transfers of 1.8\,MB segments representing a 3–4\,s adaptive bitrate video chunk. 
The QoE metric is MOS derived from segment fetch time using an IQX-based exponential model.

\end{enumerate}

\begin{figure}
    \centering
    \includegraphics[width=\columnwidth]{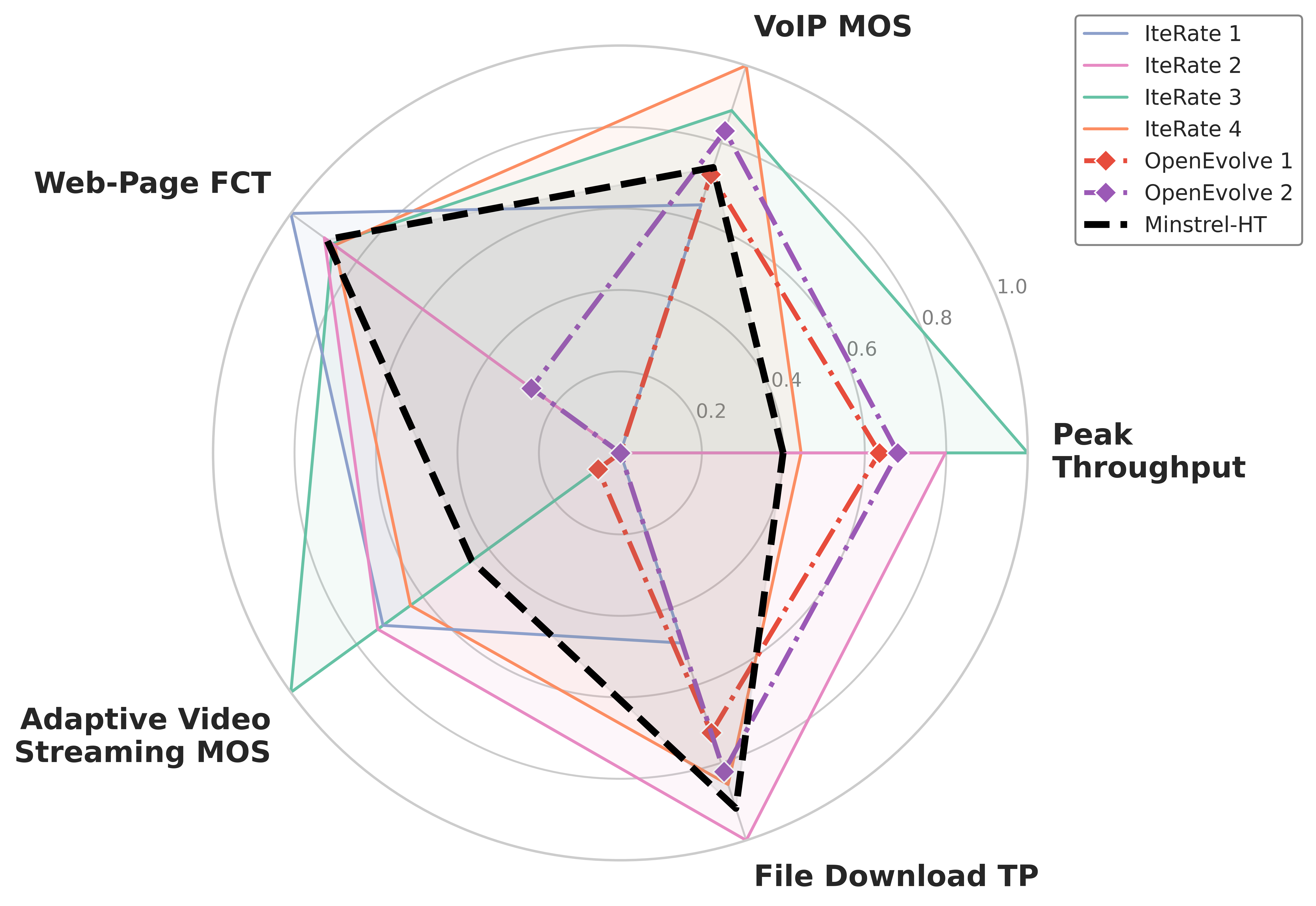}
    \caption{Radar plot comparing \name to OpenEvolve and Minstrel-HT across 5 workloads. \textnormal{Each axis corresponds to a different application workload and its associated QoE metric, with scores normalized so that higher values indicate better application-level performance. Across the five workloads, \name discovers four policies that outperform both OpenEvolve’s candidates and the Minstrel-HT baseline on their respective application metrics, illustrating the ability of \name to design good workload-specific algorithms rather than relying on a static rate-control policy.}}

\label{fig:radarize}
\end{figure}

\subsection{Baselines}
\label{subsec:baselines}
\begin{figure*}[t]

\hfill
\begin{subfigure}{0.31\textwidth}
    \centering
    \includegraphics[width=\textwidth]{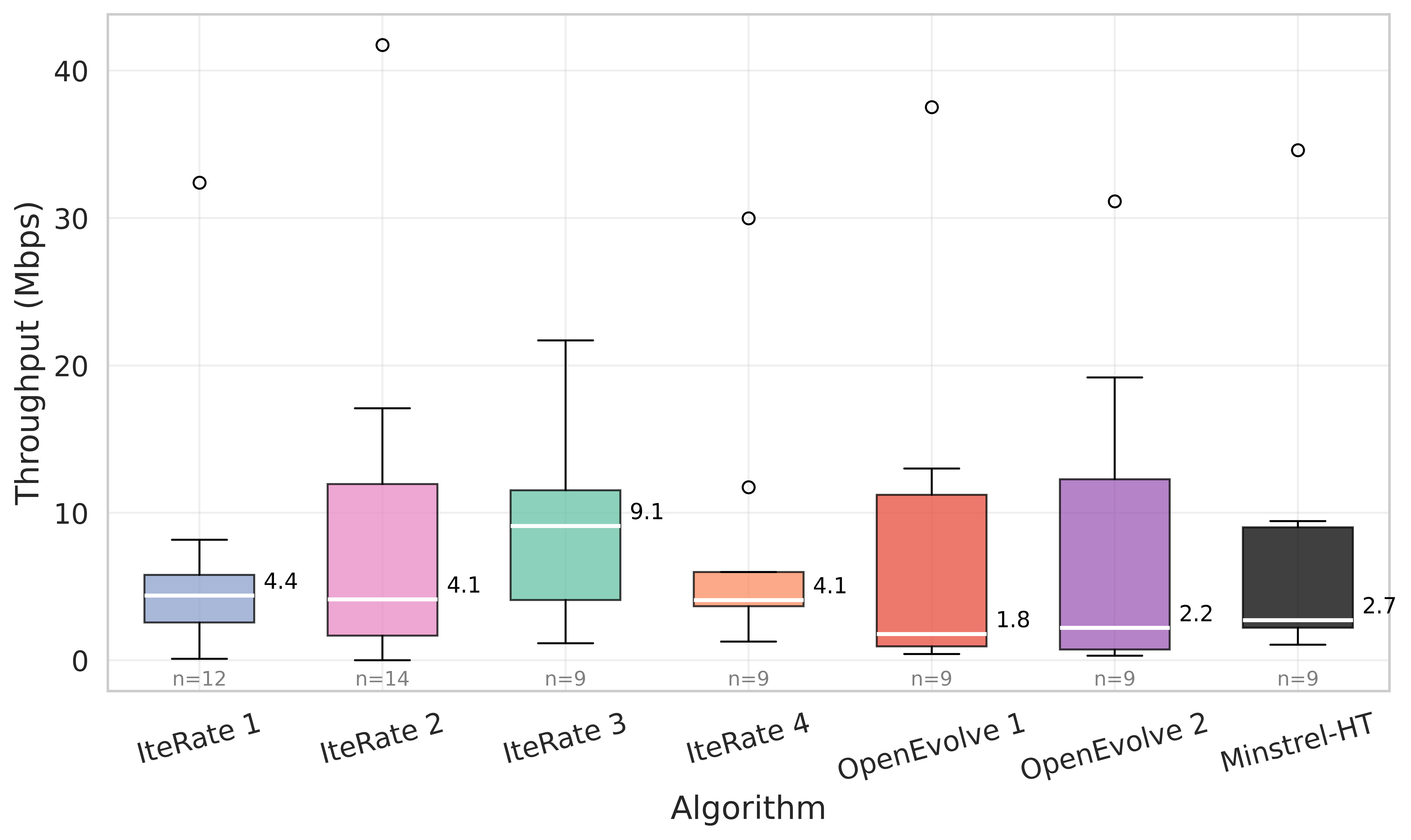}
    \caption{Peak Throughput}
    \label{fig:evala}
\end{subfigure}
\hfill
\begin{subfigure}{0.31\textwidth}
    \centering
    \includegraphics[width=\textwidth]{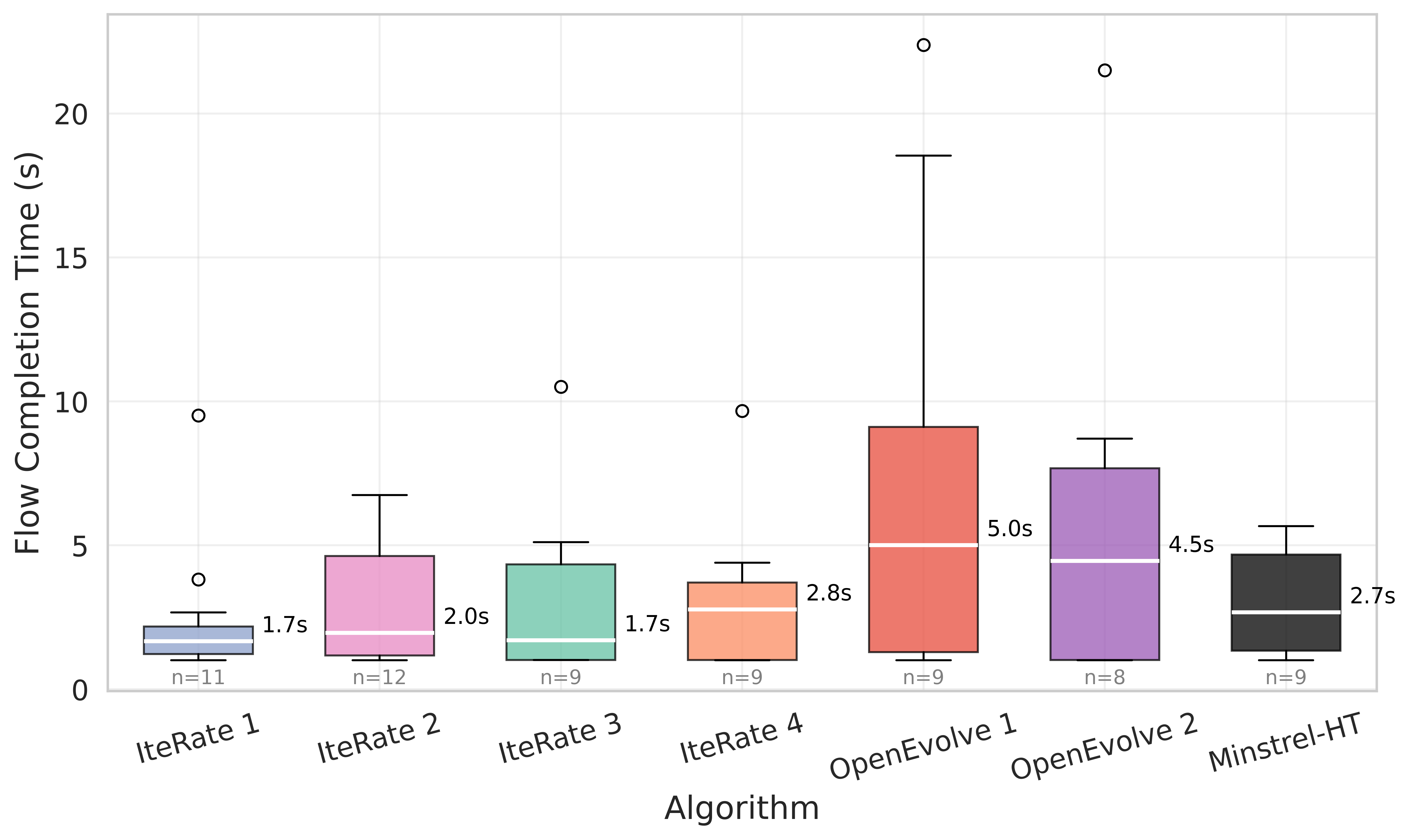}
    \caption{Web Flow Completion Time}
    \label{fig:evalb}
\end{subfigure}
\hfill
\begin{subfigure}{0.31\textwidth}
    \centering
    \includegraphics[width=\textwidth]{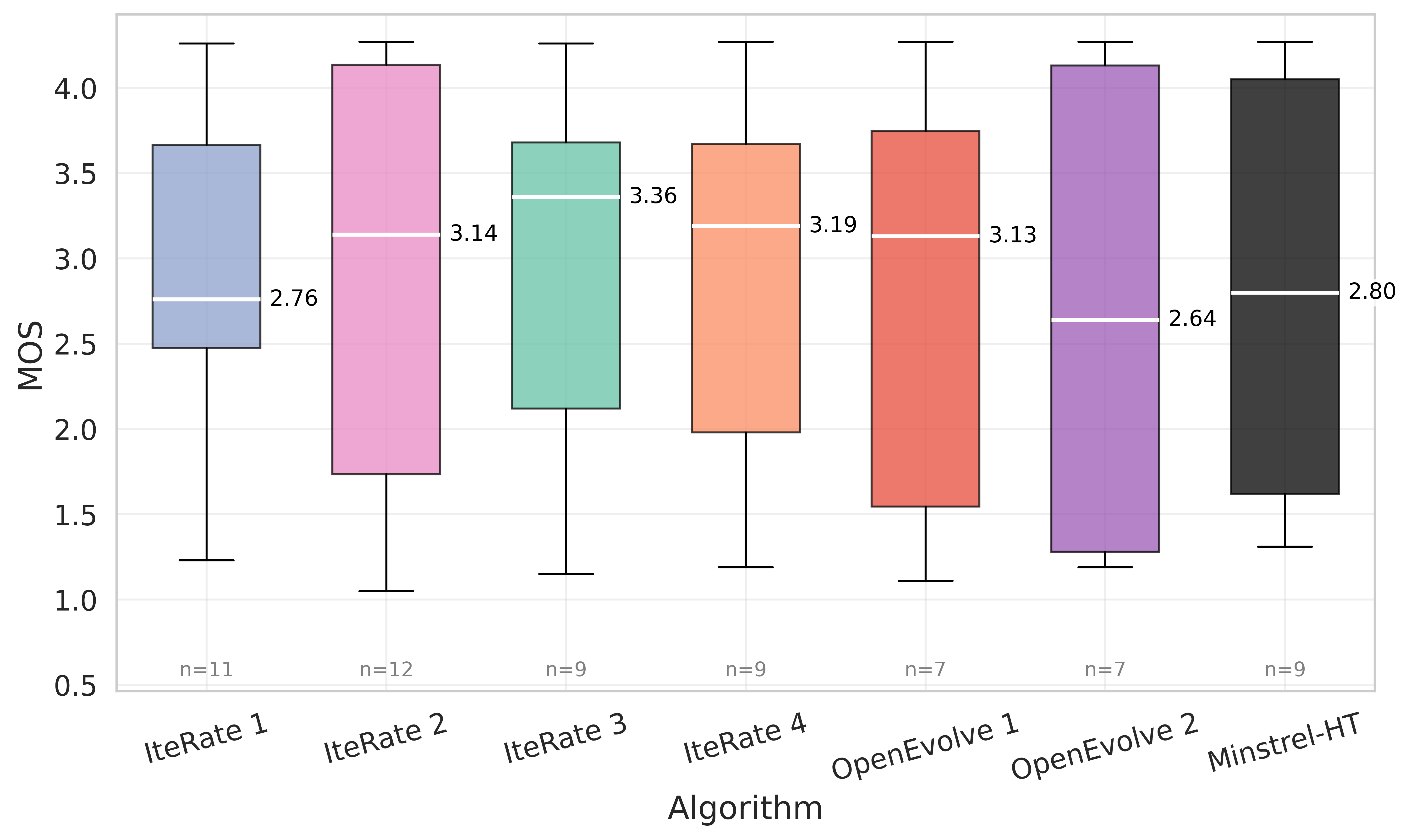}
    \caption{Adaptive Video Performance}
    \label{fig:evalc}
\end{subfigure}

\caption{Individual comparisons of each algorithm on each specific workload. \textnormal{Across heterogeneous workloads, no single algorithm consistently dominates; instead, performance varies depending on the application characteristics. Our key advantage is that the agentic system can infer, from the observed workload, which algorithm is most appropriate for the current application. As the application executes, the agent continues to explore and evaluate policies under new workload conditions, enabling ongoing adaptation.}}
\label{fig:maineval}

\end{figure*}

We compare \name against two classes of bitrate selection algorithms to evaluate its performance across heuristic, statistical, and evolutionary paradigms:

\begin{enumerate}[leftmargin=*, labelindent=0pt, itemsep=1pt]
    \item \textbf{Minstrel \cite{minstrel}:} The industry-standard rate adaptation algorithm in the Linux kernel. It uses an Exponentially Weighted Moving Average (EWMA) of packet success probabilities and dedicates 10\% of traffic to random sampling to find the best throughput.

    \item \textbf{OpenEvolve \cite{openevolve}:} A code-mutation evolutionary baseline that also interacts directly with the testbed to refine the bitrate algorithm. It executes an iterative process where the results of each testbed run are observed and used to evolve the bitrate algorithm over successive generations. We evaluate the best algorithm OpenEvolve is able to produce within a 1-hour window of evolution.

\end{enumerate}

\subsection{Performance across applications}
We plot the relative performance across five workloads for representative algorithms from three families: \textbf{IteRate} policies, \textbf{OpenEvolve}-generated policies, and the widely deployed \textbf{Minstrel-HT} baseline. Each axis corresponds to a distinct workload capturing a different application objective. Metrics are normalized per workload so that the best-performing algorithm attains a score of 1.0, and others are scaled accordingly.

The results reveal clear workload-dependent tradeoffs. For example, \textbf{IteRate 3} (Appendix \ref{app:code}) achieves the highest normalized peak throughput (1.00) and adaptive video QoE (1.00), while maintaining strong web performance (0.96). The IteRate 4 policy achieves the best VoIP quality (1.00) and strong file download throughput (0.85). In contrast, OpenEvolve-generated policies exhibit substantial imbalance: OpenEvolve 2 achieves moderate throughput (0.68) and VoIP quality (0.82), but performs poorly on latency-sensitive workloads such as web transfers (0.29) and adaptive video (0.00). The Minstrel baseline provides relatively balanced but consistently lower performance across workloads (e.g., 0.40 peak throughput and 0.44 video QoE).

Overall, the IteRate policies span a larger area across the radar axes, indicating stronger performance across diverse workloads. This shows that by specializing rate selection to application objectives, IteRate consistently outperforms Minstrel-HT and OpenEvolve across scenarios.

\label{subsec:dynamics}

\subsection{Individual Workload Analysis}
We present \name 1-4, OpenEvolve 1-2 and the Minstrel-HT baseline in terms of their performance on peak throughput, web flow completion time, and adaptive video performance. 

Figure \ref{fig:evala} reports peak throughput under saturated traffic. 
\textbf{IteRate 3} achieves the highest median throughput (9.1\,Mbps), substantially exceeding both Minstrel (2.7\,Mbps) and OpenEvolve policies (1.8--2.2\,Mbps). 
Other IteRate variants maintain competitive performance while preserving stability across links. 
This indicates that the agent can synthesize rate policies that aggressively exploit high-quality links while avoiding unstable rate selections.

Figure \ref{fig:evalb} compares web page flow completion time (FCT), where lower values are better. 
\textbf{IteRate 1} and \textbf{IteRate 3} achieve the lowest median FCT (1.7\,s), outperforming both Minstrel (2.7\,s) and OpenEvolve policies (4.5--5.0\,s). 
This improvement highlights the importance of application-aware rate control: by prioritizing reliability and short bursts of traffic, IteRate significantly reduces web latency compared to generic throughput-oriented algorithms.

Figure \ref{fig:evalc} shows the distribution of video MOS across algorithms. 
\textbf{IteRate 3} achieves the highest median MOS (3.36, a score described as Excellent), outperforming both OpenEvolve policies and the Minstrel-HT baseline. 
\textbf{IteRate 2} and \textbf{IteRate 4} also maintain strong video quality with medians around 3.1--3.2, indicating that application-aware adaptation can sustain high streaming quality across varying link conditions. 
In contrast, OpenEvolve policies exhibit lower medians (2.64--3.13), while Minstrel provides moderate but less optimized performance (2.80). 
These results demonstrate that IteRate can explicitly optimize for latency-sensitive real-time traffic.

\label{subsec:overhead}

\section{Case Study: Agentic Protocol Evolution}
\label{sec:agent-trajectory}

To understand how \name discovers a working design, we analyze an example
trajectory of its reasoning and experiments. Rather than behaving like a
black-box optimizer, \name repeatedly proposes a hypothesis, executes an evaluation protocol, diagnoses the resulting failure mode, and
refines the algorithm accordingly. In this experiment, the agent controls one sender and two receivers (R1 \& R2). 

\paragraph{Iteration 1: Change Detection Hypothesis.}
At the beginning of the problem, the Scientist tried to frame rate control as a non-stationary bandit problem. The agent proposed:

\begin{promptbox}
\small
``Piecewise-nonstationary BPF rate control with change detection will improve
over soak time\dots local exploration around the current best arm\dots
change-point detection resets.'' 
\end{promptbox}

This proposal already shows a research idea that includes some form of temporal reasoning. The agent believed that the bitrate selection algorithm should behave differently after minutes of operation than in the initial seconds. The hypothesis was operationalized through an experiment configuration that extended the Minstrel baseline.

However, the resulting experiments falsified the hypothesis. Under the application workload and another test the agent decided to run, the adaptive algorithm selected only MCS0. For R1, throughput dropped from 8.91\,Mbps to 2.88\,Mbps during the first trial, but after the agent decided to run a second trial, it observed that the performance dropped from 26.87\,Mbps to 2.66\,Mbps. Similar degradation occurred on R2. The agent summarized the result succinctly: the controller ``collapses to MCS0 and does not improve over a long time.''

Rather than adjusting parameters, the agent identified the mechanism of
failure as {\em catastrophic lock-in to the lowest rate}. The next hypothesis
explicitly targeted this failure mode, predicting that a successful design
must ``spend substantially less airtime at MCS0.''

\paragraph{Iterations 2--3: Verifier Constraints.}
The next two revisions introduced per-MCS EWMA statistics (similar to Minstrel) and deterministic retests to prevent lock-in. However, both implementations were rejected by the AP's eBPF verifier:

\begin{promptbox}
\small
``Deployment/load verification failed with \texttt{reg type unsupported
for arg\#0 function on\_tx\_status\#42}\dots appears to be a
loader/verifier compatibility issue for this program shape.''
\end{promptbox}

This discovery changed the search space. The agent inferred that only the
callback and state ``skeleton'' of the previously accepted program could
reliably pass verification. Instead of abandoning the design, the agent
promoted this observation into a new rule: all subsequent algorithms must
preserve the exact skeleton of the verified implementation.

\paragraph{Iteration 4: Skeleton-Safe Anti-Collapse.}
Using the new constraint, the agent implemented a minimal modification that added a periodic retest and held values for longer.  This version successfully eliminated collapse, but exposed a new limitation.
Telemetry showed that nearly all transmissions used MCS4 (14847 of 14848 frames), producing lower throughput than the baseline. The algorithm had solved the collapse but became overly conservative. The agent concluded that the next design must maintain the anti-collapse property while enabling local upward motion in the rate ladder.

\paragraph{Iteration 7: Local Promotion and Demotion.}
The next revision introduced small local adjustments around the current rate,
allowing promotion or demotion based on recent outcomes while preserving the
anti-collapse safeguards.

This design produced the first asymmetric result. For R1, throughput improved dramatically, increasing from 13.04\,Mbps to 21.91\,Mbps while reducing RTT from 87\,ms to 60\,ms. However, R2 regressed slightly from 13.61\,Mbps to 10.38\,Mbps. The agent described the outcome as
``partially supported,'' indicating that the mechanism worked for strong links
but remained fragile for weaker ones.

\begin{figure}
    \centering
    \includegraphics[width=\columnwidth]{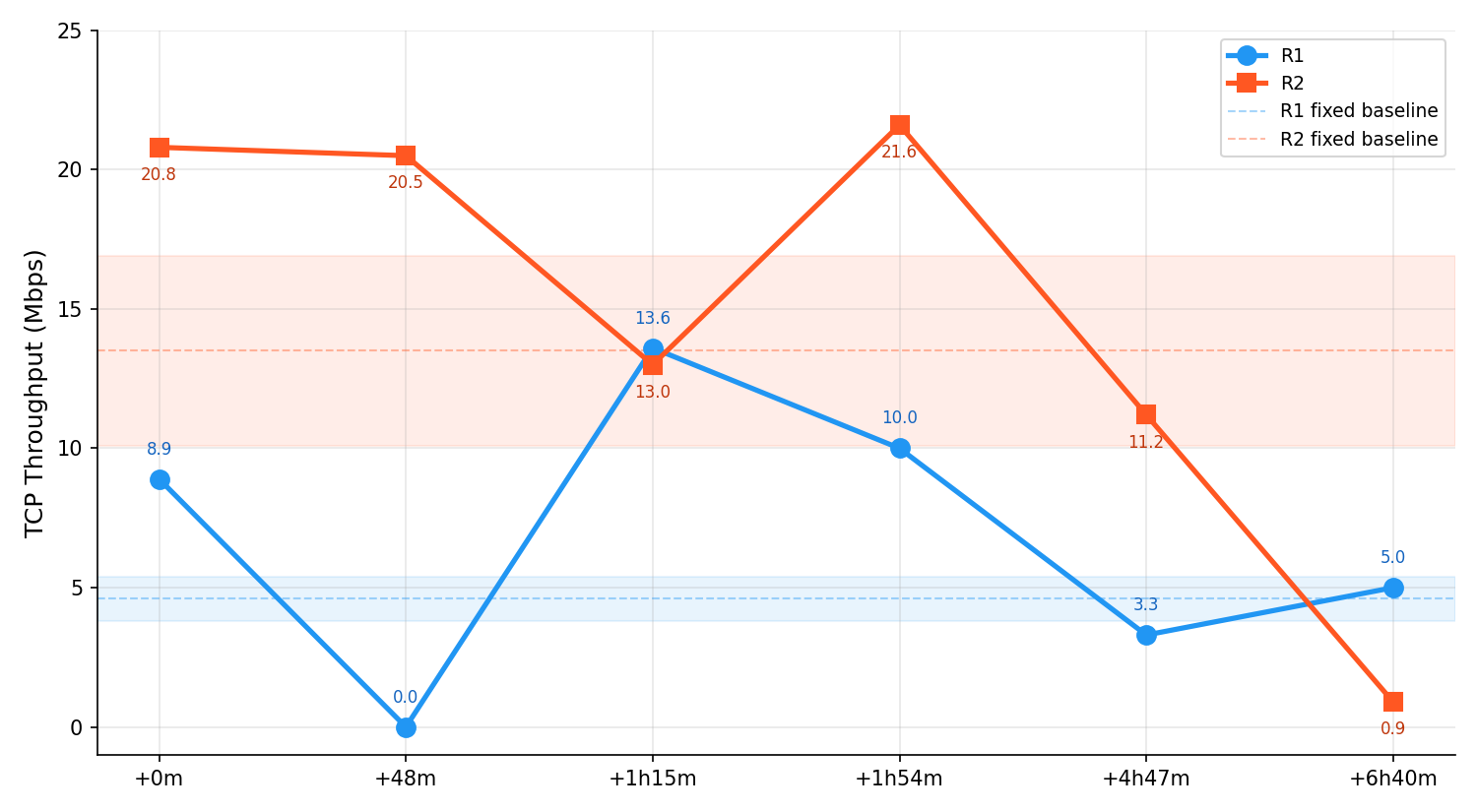}
    \caption{\name iterations over wall time. \textnormal{The agent explores successive versions (not necessarily monotonic improvements) until debugging/redeployment time exceeds a limit or progress stalls. It converges to a satisfactory algorithm in 75 minutes, even though its initial hypothesis already outperformed the naive baseline.}}
\label{fig:iterate}
\end{figure}

\paragraph{Iteration 9: Conservative Promotion.}
To address this asymmetry, \name proposed a more conservative 
policy:

\begin{promptbox}
\small
``Require repeated clean evidence before sustained promotion to MCS5 and
retreat faster after retry inflation.''
\end{promptbox}

This refinement produced the first across-the-board improvement. For R1,
throughput increased from 3.77\,Mbps to 9.98\,Mbps in the early phase
and from 6.40\,Mbps to 7.82\,Mbps during soak while reducing retry
rates substantially. R2 also improved, reaching 21.57\,Mbps and
reducing UDP loss from 2.49\% to 1.44\%.

\paragraph{Iterations 12--15: Weak-Link Hysteresis.}
Here the agent decided to test whether this approach would work across varied links. It changed the transmit power of the sender to 10\,dBm, and observed that the algorithm struggled on very weak links. The agent diagnosed the new failure mode:

\begin{promptbox}
\small
``The controller remains too willing to revisit MCS5 under persistent retry
inflation.''
\end{promptbox}

The solution introduced explicit hysteresis around MCS5:

\begin{promptbox}
\textbf{MCS5 Cooldown Logic}

\begin{tabbing}
\hspace{1.5em}\=\hspace{1.5em}\=\kill
\texttt{if current\_mcs == MCS5 \&\&} \\
\>\texttt{(tx\_failure || high\_retry): mcs5\_cooldown $\leftarrow$ COOLDOWN} \\
\texttt{if mcs5\_cooldown > 0: current\_mcs $\leftarrow$ min(current\_mcs, MCS4)} \\
\texttt{if clean\_successes $\ge$ 4: mcs5\_cooldown $\leftarrow$ 0}
\end{tabbing}

\end{promptbox}
This modification improved degraded links under normal conditions, but did not
fully recover near-outage links.

\paragraph{Iteration 27: Outage Guard.}
The final refinement targeted near-outage regimes explicitly. The new
mechanism introduced an \texttt{outage\_guard} that forces the controller to
remain at MCS3 until a sequence of clean transmissions is observed:

\begin{promptbox}
\begin{tabbing}
\hspace{1.5em}\=\hspace{1.5em}\=\kill
\texttt{if failure\_or\_high\_retry(mcs >= 4):} \\
\>\texttt{last\_good\_mcs $\leftarrow$ max(MCS3, current\_mcs - 1)} \\
\>\texttt{outage\_guard $\leftarrow$ GUARD\_HOLD} \\
\texttt{if outage\_guard > 0:} \\
\>\texttt{current\_mcs $\leftarrow$ max(MCS3, last\_good\_mcs)} \\
\>\texttt{if clean\_successes >= GUARD\_EXIT:} \\
\>\>\texttt{outage\_guard $\leftarrow$ 0}
\end{tabbing}

\end{promptbox}

The results here were promising, and the links retain the gains achieved in earlier iterations, producing \name Algorithm 3 above (Appendix \ref{app:code}). 

This trajectory reveals a structured convergence process. The agent begins with a broad theoretical idea, falsifies it through experiments, discovers deployment constraints imposed by the verifier, and progressively refines the controller to address concrete failure modes. Each stage adds a new mechanism until the algorithm stabilizes across a range of link conditions. The agent repeatedly applies a simple reasoning loop: propose a mechanism, evaluate it under a fixed protocol, identify the failure mode, and redesign the controller to address that specific weakness.
\section{Conclusion}
This paper presented \name, an autonomous research system to synthesize real-world Wi-Fi rate control algorithms. \name uses a multi-agent AI architecture to conduct the full scientific cycle: formulating hypotheses, writing eBPF programs that run inside the Linux kernel, deploying them over-the-air to Wi-Fi devices, collecting fine-grained telemetry for analysis, and iterating based on experimental evidence, all without human intervention. Performance on a 58-node testbed shows that its algorithms outperform Minstrel. 
More importantly, results show the agentic architecture can effectively automate research for non-trivial problems in wireless and networked systems.

\bibliographystyle{ACM-Reference-Format}
\bibliography{reference}

@inproceedings{lacage2004practical,
  author    = {Mathieu Lacage and Mohammad Hossein Manshaei and Thierry Turletti},
  title     = {{IEEE} 802.11 Rate Adaptation: A Practical Approach},
  booktitle = {Proceedings of the 7th ACM International Symposium on Modeling, Analysis and Simulation of Wireless and Mobile Systems (MSWiM)},
  pages     = {126--134},
  year      = {2004}
}

@mastersthesis{bicket2005bitrate,
  author = {John C. Bicket},
  title  = {Bit-Rate Selection in Wireless Networks},
  school = {Massachusetts Institute of Technology},
  year   = {2005}
}

@inproceedings{wong2006robust,
  author    = {Starsky H. Y. Wong and Songwu Lu and Hao Yang and Vaduvur Bharghavan},
  title     = {Robust Rate Adaptation for 802.11 Wireless Networks},
  booktitle = {Proceedings of the 12th Annual International Conference on Mobile Computing and Networking (MobiCom)},
  pages     = {146--157},
  year      = {2006}
}

@inproceedings{xia2013minstrel,
  author    = {Dong Xia and Jonathan Hart and Qiang Fu},
  title     = {Evaluation of the Minstrel Rate Adaptation Algorithm in {IEEE} 802.11g {WLANs}},
  booktitle = {Proceedings of the IEEE International Conference on Communications (ICC)},
  pages     = {2223--2228},
  year      = {2013}
}

@inproceedings{vutukuru2009softrate,
  author    = {Mythili Vutukuru and Hari Balakrishnan and Kyle Jamieson},
  title     = {Cross-Layer Wireless Bit Rate Adaptation},
  booktitle = {Proceedings of the ACM SIGCOMM 2009 Conference on Data Communication},
  pages     = {3--14},
  year      = {2009}
}

@inproceedings{abedi2014trate,
  author    = {Ali Abedi and Tim Brecht},
  title     = {{T-RATE}: A Framework for the Trace-Driven Evaluation of 802.11 Rate Adaptation Algorithms},
  booktitle = {Proceedings of the IEEE 22nd International Symposium on Modeling, Analysis, and Simulation of Computer and Telecommunication Systems (MASCOTS)},
  pages     = {1--10},
  year      = {2014}
}

@inproceedings{abedi2016tsimn,
  author    = {Ali Abedi and Andrew Heard and Tim Brecht},
  title     = {{T-SIMn}: Towards the High Fidelity Trace-Based Simulation of 802.11n Networks},
  booktitle = {Proceedings of the 19th ACM International Conference on Modeling, Analysis and Simulation of Wireless and Mobile Systems (MSWiM)},
  pages     = {83--92},
  year      = {2016}
}

@article{karmakar2017smartla,
  author  = {Raja Karmakar and Samiran Chattopadhyay and Sandip Chakraborty},
  title   = {{SmartLA}: Reinforcement Learning-Based Link Adaptation for High Throughput Wireless Access Networks},
  journal = {Computer Communications},
  volume  = {110},
  pages   = {1--25},
  year    = {2017}
}

@article{karmakar2020banditlink,
  author  = {Raja Karmakar and Samiran Chattopadhyay and Sandip Chakraborty},
  title   = {An Online Learning Approach for Auto Link-Configuration in {IEEE} 802.11ac Wireless Networks},
  journal = {Computer Networks},
  volume  = {181},
  pages   = {107426},
  year    = {2020}
}

@inproceedings{khastoo2020neura,
  author    = {Shervin Khastoo and Tim Brecht and Ali Abedi},
  title     = {{NeuRA}: Using Neural Networks to Improve {WiFi} Rate Adaptation},
  booktitle = {Proceedings of the 23rd ACM International Conference on Modeling, Analysis and Simulation of Wireless and Mobile Systems (MSWiM)},
  pages     = {161--170},
  year      = {2020}
}

@inproceedings{queiros2022dara,
  author    = {Ruben Queir{\'o}s and Eduardo Nuno Almeida and Helder Fontes and Jos{\'e} Ruela and Rui Campos},
  title     = {{Wi-Fi} Rate Adaptation using a Simple Deep Reinforcement Learning Approach},
  booktitle = {Proceedings of the IEEE Symposium on Computers and Communications (ISCC)},
  year      = {2022}
}

@inproceedings{smartrate,
  author    = {Malik Ahmad Yar Khan and Darryl Veitch},
  title     = {SmartRate: A new dynamic rate adaptation algorithm for 802.11 wireless networks},
  booktitle = {12th IEEE International Symposium on a World of Wireless, Mobile and Multimedia Networks (WoWMoM)},
  pages     = {1--10},
  year      = {2011},
  publisher = {IEEE Computer Society},
  doi       = {10.1109/WOWMOM.2011.5986389}
}

@article{raca,
  author    = {Tingpei Huang and Shibao Li and Shaoshu Gao},
  title     = {RaCA: A joint rate and channel adaptation scheme for dense 802.11n networks},
  journal   = {Procedia Computer Science},
  volume    = {111},
  pages     = {183--189},
  year      = {2017},
  doi       = {10.1016/j.procs.2017.06.026}
}

@article{damysus,
  author    = {Ioannis Selinis and Konstantinos Katsaros and Seiamak Vahid and Rahim Tafazolli},
  title     = {Damysus: A Practical IEEE 802.11ax BSS Color Aware Rate Control Algorithm},
  journal   = {International Journal of Wireless Information Networks},
  volume    = {26},
  number    = {4},
  pages     = {285--307},
  year      = {2019},
  doi       = {10.1007/S10776-019-00439-6}
}

@inproceedings{mudra,
  author    = {Varun Gupta and Craig Gutterman and Yigal Bejerano and Gil Zussman},
  title     = {Experimental evaluation of large scale WiFi multicast rate control},
  booktitle = {35th Annual IEEE International Conference on Computer Communications (INFOCOM)},
  pages     = {1--9},
  year      = {2016},
  publisher = {IEEE},
  doi       = {10.1109/INFOCOM.2016.7524343}
}

@misc{extreme-wing-rate-selection,
  author       = {{Extreme Networks}},
  title        = {Access Point System Reference Guide: Rate Selection Methods},
  howpublished = {\url{https://documentation.extremenetworks.com/WiNG/7.3.1/APSRG/GUID-3D70DA57-4C08-4220-A19E-3A95522CCA2C.shtml}},
  note         = {WiNG 7.3.1 documentation; accessed 2026-03-12}
}

@misc{mist-data-rates,
  author       = {{Juniper Networks}},
  title        = {Wi-Fi Data Rate Configuration | Mist},
  howpublished = {\url{https://www.juniper.net/documentation/us/en/software/mist/mist-wireless/topics/ref/mist-data-rates.html}},
  note         = {accessed 2026-03-12}
}

@misc{mist-ap-overview,
  author       = {{Juniper Networks}},
  title        = {Overview of Juniper APs | Mist},
  howpublished = {\url{https://www.juniper.net/documentation/us/en/software/mist/mist-wireless/topics/concept/mist-wireless-guide-ap-overview.html}},
  note         = {accessed 2026-03-12}
}

@misc{meraki-health,
  author       = {{Cisco Meraki}},
  title        = {Meraki Health Overview},
  year         = {2025},
  howpublished = {\url{https://documentation.meraki.com/Platform_Management/Dashboard_Administration/Operate_and_Maintain/Monitoring_and_Reporting/Meraki_Health_Overview}},
  note         = {accessed 2026-03-12}
}

@misc{cisco-ai-rrm,
  author       = {{Cisco}},
  title        = {Cisco Catalyst Center AI-Enhanced RRM Deployment Guide},
  year         = {2026},
  howpublished = {\url{https://www.cisco.com/c/en/us/td/docs/wireless/controller/9800/technical-reference/ai-enhanced-rrm-dg.html}},
  note         = {accessed 2026-03-12}
}

@misc{openevolve,
  author       = {{OpenEvolve Authors}},
  title        = {OpenEvolve},
  howpublished = {\url{https://github.com/codelion/openevolve}},
  note         = {GitHub repository; accessed 2026-03-13}
}

@misc{glia,
  author = "P. Hamadanian and P. Karimi and A. Nasr-Enfahany and K. Noorbakhsh and J. Chandler and A. Parandeh and M. Alizadeh and H. Balakrishnan",
  title        = "{Glia: A Human-Inspired AI for Automated Systems Design and Optimization}",
  howpublished = {\url{https://arxiv.org/abs/2510.27176}},
}

@article{ADRS,
  title= "{Let the Barbarians In: How AI Can Accelerate Systems Performance Research}",
  author={Cheng, Audrey and Liu, Shu and Pan, Melissa and Li, Zhifei and Agarwal, Shubham and Cemri, Mert and Wang, Bowen and Krentsel, Alexander and Xia, Tian and Park, Jongseok  },
  journal={arXiv preprint arXiv:2512.14806},
  year={2025}
}

@inproceedings{morpheus,
  title={Towards infrastructure-assisted wifi rate adaptation for converged networks with morpheus},
  author={Thapa, Prashiddha D and Kappen, Arne and Schulz-Zander, Julius},
  booktitle={Proceedings of the 19th Workshop on Mobility in the Evolving Internet Architecture},
  pages={19--24},
  year={2024}
}

@article{mira,
  author  = {Ilias Pefkianakis and Sunghyun Yun and Songwu Lu},
  title   = {MIRA: A Multi-Path MIMO Rate Adaptation Algorithm for IEEE 802.11ac},
  journal = {IEEE/ACM Transactions on Networking},
  volume  = {24},
  number  = {6},
  pages   = {3635--3648},
  year    = {2016},
  doi     = {10.1109/TNET.2015.2504984}
}

@inproceedings{samplelite,
  author    = {Lefteris Kriara and Konstantinos Katsaros and George Xylomenos and Ioannis Stavrakakis},
  title     = {SampleLite: Fast and Efficient Adaptive Rate Selection},
  booktitle = {IFIP Networking Conference (IFIP Networking)},
  year      = {2015},
  pages     = {1--9},
  doi       = {10.1109/IFIPNetworking.2015.7145308}
}

@inproceedings{ramas,
  author    = {Nguyen Le Minh and Choi Hee Yong and Choi Dongho and Kim Dongkyun and Choi Choong Seon},
  title     = {RAMAS: Rate adaptation for Mobile users in 802.11n},
  booktitle = {2011 IEEE International Conference on Communications (ICC)},
  year      = {2011},
  pages     = {1--5},
  doi       = {10.1109/icc.2011.5963415}
}

@inproceedings{cara,
  author    = {Jinseok Kim and Seongkwan Kim and Sunghyun Choi and Daji Qiao},
  title     = {CARA: Collision-Aware Rate Adaptation for IEEE 802.11 WLANs},
  booktitle = {IEEE INFOCOM 2006},
  year      = {2006},
  pages     = {1--11},
  doi       = {10.1109/INFOCOM.2006.95}
}

@inproceedings{ear,
  author    = {Glenn Judd and Xiaohui Wang and Peter Steenkiste},
  title     = {Efficient Channel-Aware Rate Adaptation in Dynamic Environments},
  booktitle = {Proceedings of the 6th International Conference on Mobile Systems, Applications, and Services (MobiSys)},
  year      = {2008},
  pages     = {118--130},
  doi       = {10.1145/1378600.1378615}
}

@inproceedings{strale,
  author    = {Hye-Sung Byeon and Sunghyun Choi and Saewoong Bahk},
  title     = {STRALE: Mobility-Aware Collaborative Rate Adaptation in Densely Deployed WLANs},
  booktitle = {IEEE INFOCOM 2017},
  year      = {2017},
  pages     = {1--9},
  doi       = {10.1109/INFOCOM.2017.8057183}
}

@article{staterate,
  author  = {Wenxuan Liu and Chao Wang and Yaguang Zhang and Yao Liu and Chunyuan Zheng and Zhiyong Feng},
  title   = {State-Aware Rate Adaptation for UAVs: A Deep Reinforcement Learning Approach},
  journal = {IEEE Transactions on Vehicular Technology},
  volume  = {69},
  number  = {10},
  pages   = {11697--11711},
  year    = {2020},
  doi     = {10.1109/TVT.2020.3017181}
}

@misc{mist-rrm,
  author       = {{Juniper Networks}},
  title        = {RRM Overview | Mist},
  howpublished = {\url{https://www.juniper.net/documentation/us/en/software/mist/mist-wireless/topics/topic-map/rrm.html}},
  note         = {accessed 2026-03-13}
}

@misc{meraki-ai-rrm,
  author       = {{Cisco Meraki}},
  title        = {Cisco Meraki Auto RF: Wi-Fi Channel and Power Management},
  howpublished = {\url{https://documentation.meraki.com/MR/Radio_Settings/Auto_RF\%3A_Wi-Fi_Channel_and_Power_Management}},
  note         = {accessed 2026-03-13}
}

@misc{aruba-airmatch,
  author       = {{HPE Aruba Networking}},
  title        = {AirMatch and ClientMatch in ArubaOS 10},
  howpublished = {\url{https://www.arubanetworks.com/techdocs/aos/wifi-design-deploy/security/multi-zone/airmatch-clientmatch/}},
  note         = {accessed 2026-03-13}
}

@misc{aruba-airmatch-brief,
  author       = {{HPE Aruba Networking}},
  title        = {How AI-Powered AirMatch Optimizes WLAN Performance},
  howpublished = {\url{https://www.hpe.com/psnow/doc/a00115435enw}},
  note         = {accessed 2026-03-13}
}

@inproceedings{minstrel,
  author    = {Dong Xia and Jonathan Hart and Qiang Fu},
  title     = {Evaluation of the Minstrel Rate Adaptation Algorithm in {IEEE} 802.11g {WLANs}},
  booktitle = {Proceedings of the IEEE International Conference on Communications (ICC)},
  pages     = {2223--2228},
  year      = {2013}
}

@inproceedings{yin2024adrx,
  author    = {Hao Yin and Murali Ramanujam and Joe Schaefer and Stan Adermann and Srihari Narlanka and Perry Lea and Ravi Netravali and Krishna Chintalapudi},
  title     = {{ADR-X}: {ANN-Assisted} Wireless Link Rate Adaptation for {Compute-Constrained} Embedded Gaming Devices},
  booktitle = {21st USENIX Symposium on Networked Systems Design and Implementation (NSDI 24)},
  year      = {2024},
  address   = {Santa Clara, CA},
  pages     = {1331--1349},
  isbn      = {978-1-939133-39-7}
}

@inproceedings{llra,
  author    = {Chi-Yu Li and Chunyi Peng and Songwu Lu and Xinbing Wang and Ranveer Chandra},
  title     = {Latency-aware rate adaptation in 802.11n home networks},
  booktitle = {2015 IEEE Conference on Computer Communications (INFOCOM)},
  year      = {2015},
  pages     = {1293--1301},
  doi       = {10.1109/INFOCOM.2015.7218505}
}

@article{woof,
  author  = {Prashanth Aravinda Kumar Acharya and Ashish Sharma and Elizabeth M. Belding and Kevin C. Almeroth and Konstantina Papagiannaki},
  title   = {Rate Adaptation in Congested Wireless Networks through Real-Time Measurements},
  journal = {IEEE Transactions on Mobile Computing},
  volume  = {9},
  number  = {11},
  pages   = {1535--1550},
  year    = {2010},
  doi     = {10.1109/TMC.2010.108}
}

@article{haratcherev2005streaming,
  author  = {Ivaylo Haratcherev and Jacco R. Taal and Koen Langendoen and Reginald L. Lagendijk and Henk J. Sips},
  title   = {Automatic IEEE 802.11 rate control for streaming applications},
  journal = {Wireless Communications and Mobile Computing},
  volume  = {5},
  number  = {4},
  pages   = {421--437},
  year    = {2005},
  doi     = {10.1002/wcm.301}
}

@inproceedings{yang2006appaware,
  author    = {Yi Yang and Mahesh K. Marina and Rajive L. Bagrodia},
  title     = {Experimental Evaluation of Application Performance with 802.11 PHY Rate Adaptation Mechanisms in Diverse Environments},
  booktitle = {IEEE Wireless Communications and Networking Conference (WCNC)},
  year      = {2006},
  pages     = {2273--2278},
  doi       = {10.1109/WCNC.2006.1696649}
}

@article{choudhury2007multimedia,
  author  = {Sayantan Choudhury and Jerry D. Gibson},
  title   = {Payload Length and Rate Adaptation for Multimedia Communications in Wireless LANs},
  journal = {IEEE Journal on Selected Areas in Communications},
  volume  = {25},
  number  = {4},
  pages   = {796--807},
  year    = {2007},
  doi     = {10.1109/JSAC.2007.361909}
}

@inproceedings{braskich2005vowlan,
  author    = {Tony Braskich and Nattavut Smavatkul and Steve Emeott},
  title     = {Optimization of a link adaptation algorithm for voice over wireless LAN applications},
  booktitle = {IEEE Wireless Communications and Networking Conference (WCNC)},
  year      = {2005},
  pages     = {1602--1607},
  doi       = {10.1109/WCNC.2005.1424753}
}

@inproceedings{choi2008voipra,
  author    = {Youngkyu Choi and Sunghyun Choi},
  title     = {A joint design of admission control and transmission rate adaptation for {VoIP} over wireless network},
  booktitle = {2008 International Symposium on a World of Wireless, Mobile and Multimedia Networks (WoWMoM)},
  year      = {2008},
  pages     = {1--12},
  doi       = {10.1109/WOWMOM.2008.4594817}
}

@article{lee2014vowlan,
  author  = {Hyewon Lee and Seongho Byeon and Byoungjin Kim and Kwang Bok Lee and Sunghyun Choi},
  title   = {Enhancing Voice over {WLAN} via Rate Adaptation and Retry Scheduling},
  journal = {IEEE Transactions on Mobile Computing},
  volume  = {13},
  number  = {12},
  pages   = {2791--2805},
  year    = {2014},
  doi     = {10.1109/TMC.2013.54}
}

@inproceedings{alomar2023causalsim,
  title={$\{$CausalSim$\}$: A causal framework for unbiased $\{$Trace-Driven$\}$ simulation},
  author={Alomar, Abdullah and Hamadanian, Pouya and Nasr-Esfahany, Arash and Agarwal, Anish and Alizadeh, Mohammad and Shah, Devavrat},
  booktitle={20th USENIX Symposium on Networked Systems Design and Implementation (NSDI 23)},
  pages={1115--1147},
  year={2023}
}

@misc{agent_drift,
      title={Technical Report: Evaluating Goal Drift in Language Model Agents}, 
      author={Rauno Arike and Elizabeth Donoway and Henning Bartsch and Marius Hobbhahn},
      year={2025},
      eprint={2505.02709},
      archivePrefix={arXiv},
      primaryClass={cs.AI},
      url={https://arxiv.org/abs/2505.02709}, 
}

@misc{mami_trafic,
  author       = {{MAMI Project}},
  title        = {trafic: Traffic Mix Generator for Network Experiments},
  year         = {2018},
  howpublished = {\url{https://github.com/mami-project/trafic}},
  note         = {GitHub repository}
}

\onecolumn
\appendix
\section{Bitrate Selection Agent --- CLI Context}
\label{app:agent}

\subsection{Overview}

The \texttt{bitrate-agent} project is a DeepAgents workflow for developing and
evaluating Wi-Fi bitrate selection (rate adaptation) algorithms on deployed
hardware. The CLI agent has direct access to device control, experiment
orchestration, ML training, and analysis tools through a persistent MQTT
connection established at session startup.

All tool calls share this connection; no subprocess or shell invocation is
required for device interaction.

\subsection{System Architecture}

The agent operates with centralized orchestration over distributed Wi-Fi
devices. Devices execute injected policies and stream telemetry, while all
algorithm design, evaluation, and learning occur centrally.

\subsubsection{Available Tool Categories}

\begin{table}[h]
\centering
\small
\begin{tabular}{p{3cm} p{4cm} p{5cm}}
\toprule
\textbf{Category} & \textbf{Tools} & \textbf{Description} \\
\midrule
MQTT Device Control & \texttt{send\_mqtt\_command}, \texttt{set\_policy}, \texttt{set\_rate}, \texttt{enable\_telemetry}, etc. & Direct device control via MQTT \\
Network Config & \texttt{configure\_wifi}, \texttt{set\_uci\_config}, etc. & OpenWRT UCI configuration \\
Experiments & \texttt{run\_iperf\_test}, \texttt{sweep\_all\_rates}, etc. & Structured experiment execution \\
Analysis & \texttt{analyze\_telemetry} & Telemetry processing \\
ML Training & \texttt{train\_pytorch\_model}, \texttt{train\_sklearn\_model} & Model lifecycle management \\
Data/Results & \texttt{list\_experiments}, \texttt{get\_best\_model} & Result queries \\
Algorithms & \texttt{deploy\_algorithm} & Algorithm management \\
Plotting & \texttt{plot\_delivery\_ratios} & Visualization \\
Workflows & \texttt{start\_workflow}, \texttt{complete\_stage} & Multi-stage pipelines \\
Search & \texttt{tavily\_search}, \texttt{arxiv\_search} & Literature search \\
Filesystem & \texttt{read\_file}, \texttt{write\_file} & File operations \\
Shell & \texttt{execute} & Shell commands \\
\bottomrule
\end{tabular}
\caption{Available CLI tool categories.}
\end{table}

\subsection{Subagent Decomposition}

Six specialized subagents are available through the \texttt{task()} interface:

\begin{itemize}
    \item \textbf{experiment-runner} — structured device experimentation
    \item \textbf{data-analyst} — telemetry analysis
    \item \textbf{ml-engineer} — model training and comparison
    \item \textbf{algorithm-designer} — rate selection logic design
    \item \textbf{evaluator} — rigorous A/B testing
    \item \textbf{network-engineer} — OpenWRT configuration
\end{itemize}

\paragraph{Delegation Rule.}
Device configuration, traffic generation, telemetry collection, and experiment
execution must be delegated via \texttt{task()}. Only read-only status queries
may be called directly.

Subagents are stateless; all relevant context (device IDs, roles, configuration,
procedure, expected outputs) must be included in each task description.

\subsection{Workflow Pipelines}

Predefined multi-stage workflows include:

\begin{itemize}
    \item \textbf{algo-dev}: baseline → analyze → design → evaluate → report
    \item \textbf{model-training}: collect → train → evaluate → report
    \item \textbf{device-characterization}: health → MCS sweep → throughput sweep → analyze
    \item \textbf{ampdu-comparison}: configuration → iperf (on/off) → analyze
\end{itemize}

Workflow pattern:
\[
\texttt{start\_workflow} \rightarrow \text{delegate} \rightarrow \texttt{complete\_stage} \rightarrow \texttt{approve\_stage}
\]

\subsection{Configuration Format}

\begin{verbatim}
{
 "broker": { "host": "...", "port": 1883, "tls": false },
 "devices": [
  { "device_id": "pineapple-abc", "role": "ap" },
  { "device_id": "pineapple-def", "role": "sta" }
 ]
}
\end{verbatim}

\subsection{Domain Constraints}

\begin{itemize}
    \item Hardware: 2.4 GHz MT76x8 (WiFi Pineapple VII)
    \item Valid MCS range: 0--7 (HT, single stream)
    \item Use HT20 or HT40 only
    \item No background traffic
    \item Must run \texttt{iperf3} before collecting telemetry
    \item Only one iperf3/ping test at a time
\end{itemize}

\subsection{Algorithm Interface}

Each algorithm in \texttt{algorithms/} implements:

\begin{verbatim}
def select_rates(
  station_telemetry: dict[int, list[dict]],
  station_macs: dict[int, str],
  current_stats: dict,
) -> dict[str, str]:
\end{verbatim}

Workflow: edit → deploy → observe telemetry → iterate.

\subsection{Safety Rules}

\begin{itemize}
    \item NEVER modify \texttt{lan} or \texttt{wan} interfaces
    \item NEVER set a gateway on ephemeral test interfaces
    \item Always assign static IPs
    \item Avoid university-prefixed SSIDs
    \item Validate node health before experiments
\end{itemize}

\subsection{Methodological Guidelines}

\begin{enumerate}
    \item Check prior experiments before running new ones
    \item Measure baseline performance first
    \item Use structured experiments
    \item Track and log all results
    \item Delegate multi-step tasks to subagents
\end{enumerate}

\subsection{Memory Section}

The system maintains a persistent memory of:

\begin{itemize}
    \item Experiment findings
    \item Model performance
    \item Device-specific observations
    \item Successful configurations
\end{itemize}

This supports iterative improvement and reproducible experimentation.

\section{Per-frame TX flags}

\begin{table}[th]
\small
\centering
\begin{tabular}{lp{4.8cm}}
\toprule
\textbf{Field} & \textbf{Description} \\
\midrule
\texttt{wcid} & Station identifier (1--127) \\
\texttt{success} & 1 if frame was acknowledged \\
\texttt{mcs\_used} & Configured MCS index \\
\texttt{retry\_count} & Number of hardware retries \\
\texttt{ewma\_per} & EWMA packet error rate \\
\texttt{tx\_\{total, success, retries\}} & Cumulative per-station counters \\
\texttt{signal} & Last RX signal strength (dBm) \\
\texttt{ack\_signal} & Last ACK signal strength (dBm) \\
\texttt{frame\_length} & SKB length in bytes \\
\texttt{timestamp\_ns} & Monotonic kernel timestamp \\
\texttt{hw\_mcs\_used} & Hardware-reported MCS\\
\texttt{is\_aggregate} & 1 if A-MPDU aggregate completion \\
\texttt{hw\_rate\_flags} & Raw HW flags (0x08\,=\,HT, 0x00\,=\,legacy) \\
\bottomrule
\end{tabular}
\caption{Per-frame TX context passed to eBPF programs. \textnormal{Fields 11--13 expose hardware-level transmission details unavailable through standard mac80211 rate control interfaces.}}
\label{tab:bpf-ctx}
\end{table}


\lstdefinestyle{bpfstyle}{
  language=C,
  basicstyle=\ttfamily\scriptsize,
  keywordstyle=\bfseries\color{blue!70!black},
  commentstyle=\itshape\color{green!50!black},
  stringstyle=\color{red!60!black},
  numberstyle=\tiny\color{gray},
  numbers=left,
  numbersep=5pt,
  xleftmargin=2em,
  frame=single,
  framesep=3pt,
  rulecolor=\color{black!30},
  breaklines=true,
  breakatwhitespace=false,
  showstringspaces=false,
  tabsize=4,
  morekeywords={__u8,__u32,__u64,__uint,__type,SEC,BPF_MAP_TYPE_ARRAY,BPF_ANY,
                __always_inline,struct,void,int,char},
  columns=fullflexible,
  keepspaces=true,
  aboveskip=0.5em,
  belowskip=0.5em,
}

\section{IteRate-generated Code}
\label{app:code}
\begin{lstlisting}[style=bpfstyle, caption={IteRate~3 (\texttt{online\_adapt\_v8.bpf.c}). Per-frame rate selection with anti-collapse memory, MCS\,5 cooldown, and near-outage guard. Invoked on every \texttt{tx\_status} callback by the kernel module.}, label={lst:iterate3}]
#define MCS_COUNT              8
#define DEFAULT_MCS            4   /* nominal operating MCS */
#define DEFAULT_LAST_GOOD      3   /* floor for last_good_mcs memory */
#define RETEST_PERIOD_MASK     15  /* re-probe every 16 frames on failure */
#define HIGH_RETRY_THRESH      2
#define VERY_HIGH_RETRY        3
#define PROMOTE_STREAK_REQ     4   /* clean frames before MCS 4 -> 5 */
#define MCS5_COOLDOWN_INIT     6
#define MID_COOLDOWN_REDUCE    2
#define OUTAGE_GUARD_INIT      10  /* frames to hold MCS 3 after outage */
#define OUTAGE_EXIT_STREAK_REQ 3   /* clean MCS 3 frames to exit guard */

struct algo_state {
    __u8  current_mcs;
    __u8  last_good_mcs;      /* highest MCS that succeeded recently */
    __u8  recent_ok;          /* any success since last failure? */
    __u8  promote_streak;     /* consecutive clean frames at DEFAULT_MCS */
    __u8  mcs5_cooldown;      /* suppress MCS 5 for N frames */
    __u8  outage_guard;       /* near-outage hold counter */
    __u8  low_ok_streak;      /* consecutive clean MCS 3 frames */
    __u8  _pad;
    __u32 frame_count;
};

SEC("syscall")
int on_tx_status(void *ctx)
{
    __u64 *args = (__u64 *)ctx;
    __u32 wcid        = (__u32)args[0];
    __u64 success     = args[1];
    __u64 mcs_used64  = args[2];
    __u64 retry_count = args[3];

    if (wcid == 0 || wcid >= BPF_MAX_STA)
        return 0;

    struct algo_state *st = bpf_map_lookup_elem(&algo_map, &wcid);
    if (!st) return 0;

    __u8 cur       = clamp(st->current_mcs, MCS_COUNT);
    __u8 last_good = clamp(st->last_good_mcs, MCS_COUNT);
    __u8 used      = clamp((__u8)mcs_used64, MCS_COUNT);

    __u8 recent_ok      = st->recent_ok;
    __u8 promote_streak = st->promote_streak;
    __u8 mcs5_cooldown  = st->mcs5_cooldown;
    __u8 outage_guard   = st->outage_guard;
    __u8 low_ok_streak  = st->low_ok_streak;
    __u32 frames        = st->frame_count + 1;
    __u8 chosen         = DEFAULT_MCS;

    /* --- Success path --- */
    if (success) {
        recent_ok = 1;
        if (used >= DEFAULT_LAST_GOOD && retry_count <= 1)
            last_good = used;           /* remember highest clean MCS */
        chosen = max(last_good, DEFAULT_MCS);
        if (chosen > 5) chosen = 5;     /* cap at MCS 5 */

        if (chosen == DEFAULT_MCS) {
            /* At MCS 4: count streak toward promotion to MCS 5 */
            promote_streak = (retry_count == 0)
                ? min(promote_streak + 1, 255) : 0;
            if (mcs5_cooldown > 0 && retry_count == 0)
                mcs5_cooldown--;
            if (mcs5_cooldown == 0
                    && promote_streak >= PROMOTE_STREAK_REQ)
                chosen = 5;             /* promote to MCS 5 */
        } else {
            if (retry_count > 0) promote_streak = 0;
            if (used >= 5 && retry_count >= 1)
                mcs5_cooldown = MID_COOLDOWN_REDUCE;
        }

    /* --- Failure path --- */
    } else {
        recent_ok = 0;
        promote_streak = 0;
        chosen = last_good;
        if (used >= 5) {
            chosen = DEFAULT_MCS;       /* drop from MCS 5 to MCS 4 */
            mcs5_cooldown = MCS5_COOLDOWN_INIT;
        } else if (used > 0 && used <= chosen) {
            chosen = used - 1;          /* step down one MCS */
        }
        chosen = max(chosen, DEFAULT_LAST_GOOD);
    }

    /* --- High-retry override --- */
    if (retry_count >= VERY_HIGH_RETRY) {
        chosen = DEFAULT_LAST_GOOD;     /* emergency drop to MCS 3 */
        promote_streak = 0;
    } else if (retry_count >= HIGH_RETRY_THRESH && chosen > DEFAULT_MCS) {
        chosen = DEFAULT_MCS;
        promote_streak = 0;
    }

    /* --- Near-outage guard --- */
    if (used >= DEFAULT_MCS
            && (!success || retry_count >= VERY_HIGH_RETRY)) {
        outage_guard = OUTAGE_GUARD_INIT;
        low_ok_streak = 0;
    } else if (outage_guard > 0 && success
            && used <= DEFAULT_LAST_GOOD && retry_count == 0) {
        low_ok_streak = min(low_ok_streak + 1, 255);
        outage_guard--;
    } else if (!success || retry_count > 0) {
        low_ok_streak = 0;
    }

    /* Periodic re-probe on sustained failure */
    if ((frames & RETEST_PERIOD_MASK) == 0 && !recent_ok)
        chosen = last_good;

    /* Enforce cooldown and outage constraints */
    if (mcs5_cooldown > 0 && chosen >= 5)
        chosen = DEFAULT_MCS;
    if (outage_guard > 0) {
        chosen = DEFAULT_LAST_GOOD;
        promote_streak = 0;
    } else if (low_ok_streak < OUTAGE_EXIT_STREAK_REQ
            && chosen > DEFAULT_LAST_GOOD) {
        chosen = DEFAULT_LAST_GOOD;
    }

    chosen = max(chosen, DEFAULT_LAST_GOOD);  /* absolute floor */

    /* Persist state and write rate to BPF map */
    st->frame_count    = frames;
    st->current_mcs    = chosen;
    st->last_good_mcs  = last_good;
    st->recent_ok      = recent_ok;
    st->promote_streak = promote_streak;
    st->mcs5_cooldown  = mcs5_cooldown;
    st->outage_guard   = outage_guard;
    st->low_ok_streak  = low_ok_streak;

    write_rate(wcid, chosen);
    return 0;
}
\end{lstlisting}

\section{Access Point}

\begin{figure}[h]
    \centering
    \includegraphics[width=0.25\linewidth]{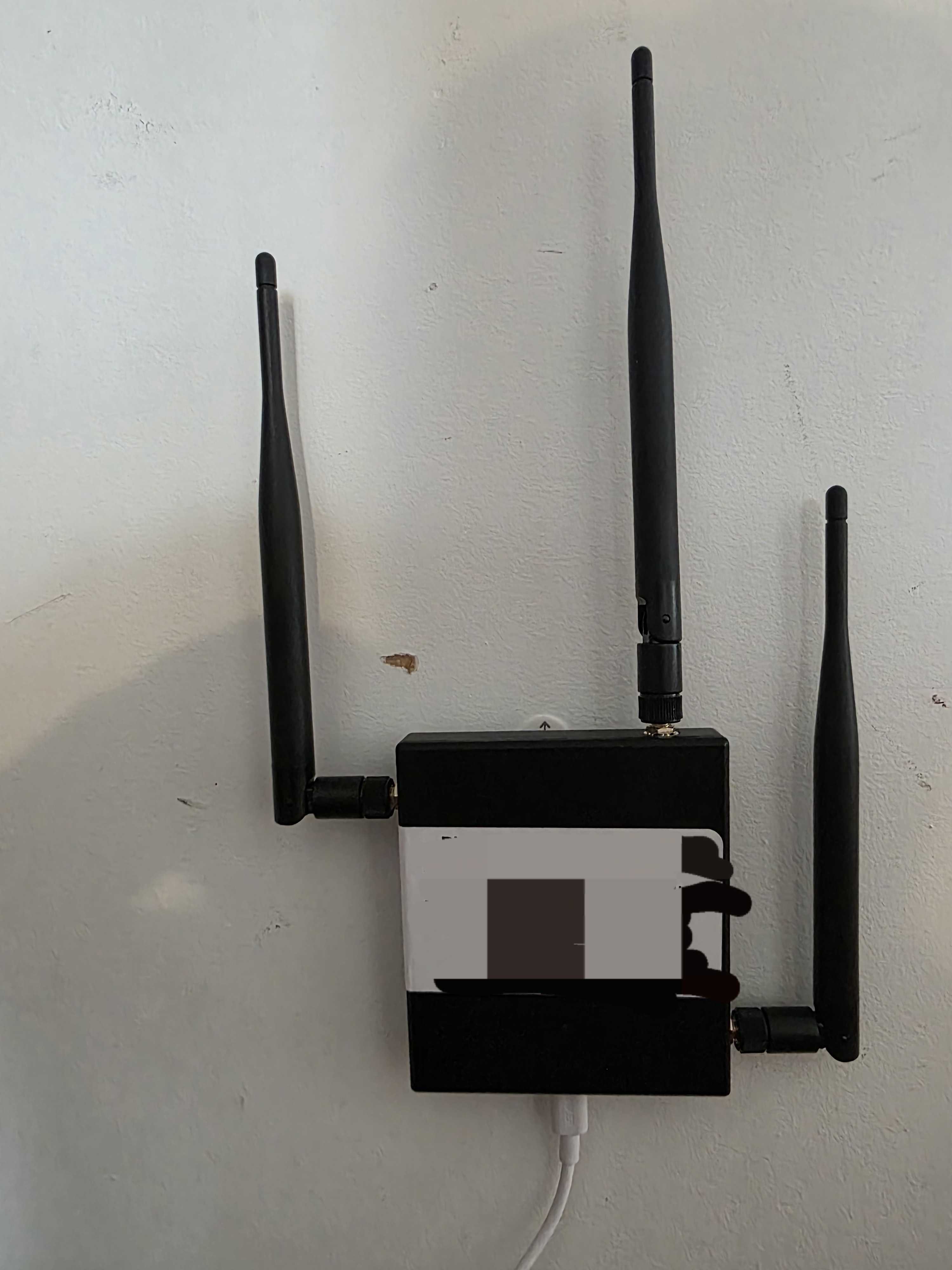}
    \caption{demo access points}
    \label{fig:placeholder}
\end{figure}

\end{document}